\documentclass[a4paper,fleqn,usenatbib]{mnras}

\usepackage{newtxtext,newtxmath}

\usepackage[T1]{fontenc}
\usepackage{ae,aecompl}

\usepackage{graphicx}	
\usepackage{amsmath}	
\usepackage{amssymb}	

\title[Clumpy dust rings]{Clumpy dust rings around non-accreting young stars}

\author[Scholz et al.]{
Aleks Scholz$^{1}$\thanks{E-mail: as110@st-andrews.ac.uk}, 
Antonella Natta$^{2}$, 
Inna Bozhinova$^{1}$, 
Maya Petkova$^{1,4}$, 
\newauthor
Howard Relles$^{1}$, 
Jochen Eisl{\"o}ffel$^{3}$
\\
$^{1}$ SUPA, School of Physics \& Astronomy, University of St Andrews, North Haugh, St Andrews, KY169SS, United Kingdom\\
$^{2}$ Dublin Institute for Advanced Studies, 31 Fitzwilliam Place, Dublin 2, Ireland\\
$^{3}$ Th{\"u}ringer Landessternwarte Tautenburg, Sternwarte 5, D-07778 Tautenburg, Germany\\
$^{4}$ Astronomisches Rechen-Institut, Zentrum f{\"u}r Astronomie der Universit{\"a}t Heidelberg, M{\"o}nchhofstrasse 12-14, 69120 Heidelberg, Germany
}

\date{Accepted XXX. Received YYY; in original form ZZZ}

\pubyear{2015}

\begin{document}
\label{firstpage}
\pagerange{\pageref{firstpage}--\pageref{lastpage}}
\maketitle

\begin{abstract}
We investigate four young, but non-accreting, very low mass stars in Orion, which show irregular eclipses by circumstellar dust. The eclipses are not recurring periodically, are variable in depth, lack a flat bottom, and their duration is comparable to the typical timescale between eclipses. The dimming is associated with reddening consistent with dust extinction. Taken together this implies the presence of rings around these four stars, with radii ranging from 0.01 to 40\,AU, comprised of optically thin dust clouds.  The stars also show IR excess indicating the presence of evolved circumstellar disks, with orders of magnitude more material than needed for the eclipses. However, the rings need to cover an opening angle of about 20 degrees to explain how common these variable stars are in the coeval population in the same region, which is more extended than a typical disk. Thus, we propose that the rings may not be part of the disks, but instead separate structures with larger scale heights. To be sustained over years, the rings need to be replenished by dust from the disk or gravitationally bound to an object (e.g., planets or planetesimals). These four stars belong to a growing and diverse class of post-T Tauri stars with dips or eclipses in their lightcurves. Dusty rings with scale heights exceeding those of disks may be a common phenomenon at stellar ages between 5 and 10\,Myr, in the transition from accretion disks to debris disks. These structures could be caused by migrating planets and may be signposts for the presence of young planetary systems. 
\end{abstract}




\section{Introduction}

A typical young star (a T Tauri star) is embedded in a complex environment extending over scales of AUs, including a gas-rich disk, accretion streams from the disk to the star, and, in the very first stages, a protostellar envelope \citep{2007prpl.conf..523W,2007prpl.conf..539M}. The interaction between a star and its disk, obscurations of the star by dust in its environment, as well as strong magnetic activity cause newly born stars to show a bewildering array of variability phenomena \citep{2014AJ....147...82C} including for example the well-known prototypes FU Ori, UX Ori, AA Tau. Once the envelope is dissipated and the disk accreted or turned into planets, the variability of young stars simplifies and is mostly constrained to spot modulations and flares, both induced by magnetic activity \citep{2007prpl.conf..297H}. 

This is the default narrative used to explain the broad pattern in the evolution of variability in young stars. But there are relevant exceptions to this timeline, stars with ages of $>5$\,Myr that maintain strong variability driven by circumstellar matter, after accretion has ceased and the mass in the disk has been strongly diminished compared to earlier stages. Typically, these stars show signs of brightness dips or eclipses induced by dusty clouds or asteroidal rings or rings around substellar companions. In all cases, the circumstellar dust has to be located along the line of sight to cause variability. Some of these variable stars still harbour evolved versions of the primordial disks typically found in much younger stars (class II objects), others host secondary dust disks presumably generated by collisions of planetesimals \citep{2007prpl.conf..573M}, or have no detectable circumstellar dust whatsoever. In any case, these objects represent excellent opportunities to study the distribution of circumstellar dust at a time when the formation of planets or planetesimals is already in an advanced stage. As these post-accretion dipping variables do not necessarily reside anymore in dense clusters, they are difficult to find -- consequently, many of them are recent and serendipitous discoveries. Deep large-scale time-domain surveys like LSST \citep{2008arXiv0805.2366I} are likely to vastly increase their numbers in the near future.

A prominent example in this category is the $>10$\,Myr old low-mass star RZ Psc, long believed to be akin to more massive and significantly younger stars named after the prototype UX Ori \citep{2010A&A...524A...8G}. Similar to UX Ori, RZ Psc is affected by eclipses caused by material along the line of sight. Its disk, however, might already be in transition to the debris stage \citep{2017RSOS....460652K}. Another example is WASP J1407, a member of the Sco-Cen association that shows recurring eclipses explained convincingly by transits of a ring system around a companion \citep{2012AJ....143...72M,2015ApJ...800..126K}. PDS 110 might be a similar type of system \citep{2017MNRAS.471..740O}, but in contrast to all other stars mentioned here it is still weakly accreting. The Kepler/K2 lightcurves covering the 5-15\,Myr old association Upper Scorpius have been a fruitful hunting ground for unusual variability in young stars. Some of the AA-tau like 'dippers' found in this region have depleted disks and lack accretion \citep{2016ApJ...816...69A}. Others defy the typical phenomenology of disk eclipses observed in younger stars and have either no measurable or only very weak disk emission 
\citep{2017ApJ...835..168D,2017AJ....153..152S}. The older cluster Pleiades seems to host a few similar objects as well \citep{2016AJ....152..113R}. The enigmatic object KIC8462852, or Boyajian's star, may indicate that dips by circumstellar matter can also occur on much older main-sequence stars \citep{2016MNRAS.457.3988B}. Recently, a brown dwarf in the $\sigma$\,Orionis cluster has been found to show eclipse-like variability, although it does not have the mid-infrared excess typical for primordial accretion disks \citep{2017arXiv170803711E}. 

In this paper we focus on four stars in Orion that are similar in age to most aforementioned objects and also show clear signs of eclipses by circumstellar material, although accretion has ceased. Originally discovered more than a decade ago, their status as non-accreting post-T Tauri stars was only confirmed recently \citep{2016MNRAS.458.3118B}. Here we present for the first time multi-wavelength lightcurves that cover timescales from days to months, combined with a detailed analysis of their spectral energy distributions and physical characteristics. We conclude that each of these four stars is surrounded by an inhomogenuous, optically thin, dusty ring that causes irregular eclipses on timescales of hours to days.

\section{Targets}
\label{targets}

The four highly variable stars discussed here have been discovered in optical lightcurves from 2001, published by \citet{2005A&A...429.1007S}, hereafter SE2005. All four are low-mass stars and part of the young population around the bright star $\epsilon$\,Ori (Alnilam), the middle belt star in Orion. All four show non-periodic variability with amplitudes unusually large for objects in this age and mass domain, and a lightcurve morphology indicating irregular eclipses. For three of them, the brightness changes occur over typical timescales of hours to days in a relatively smooth manner, whereas the fourth (V2227) varies much more rapidly over timespans of minutes. Their lightcurves from 2001 are reproduced in Fig. 1.  

\begin{figure*}
\includegraphics[width=16cm]{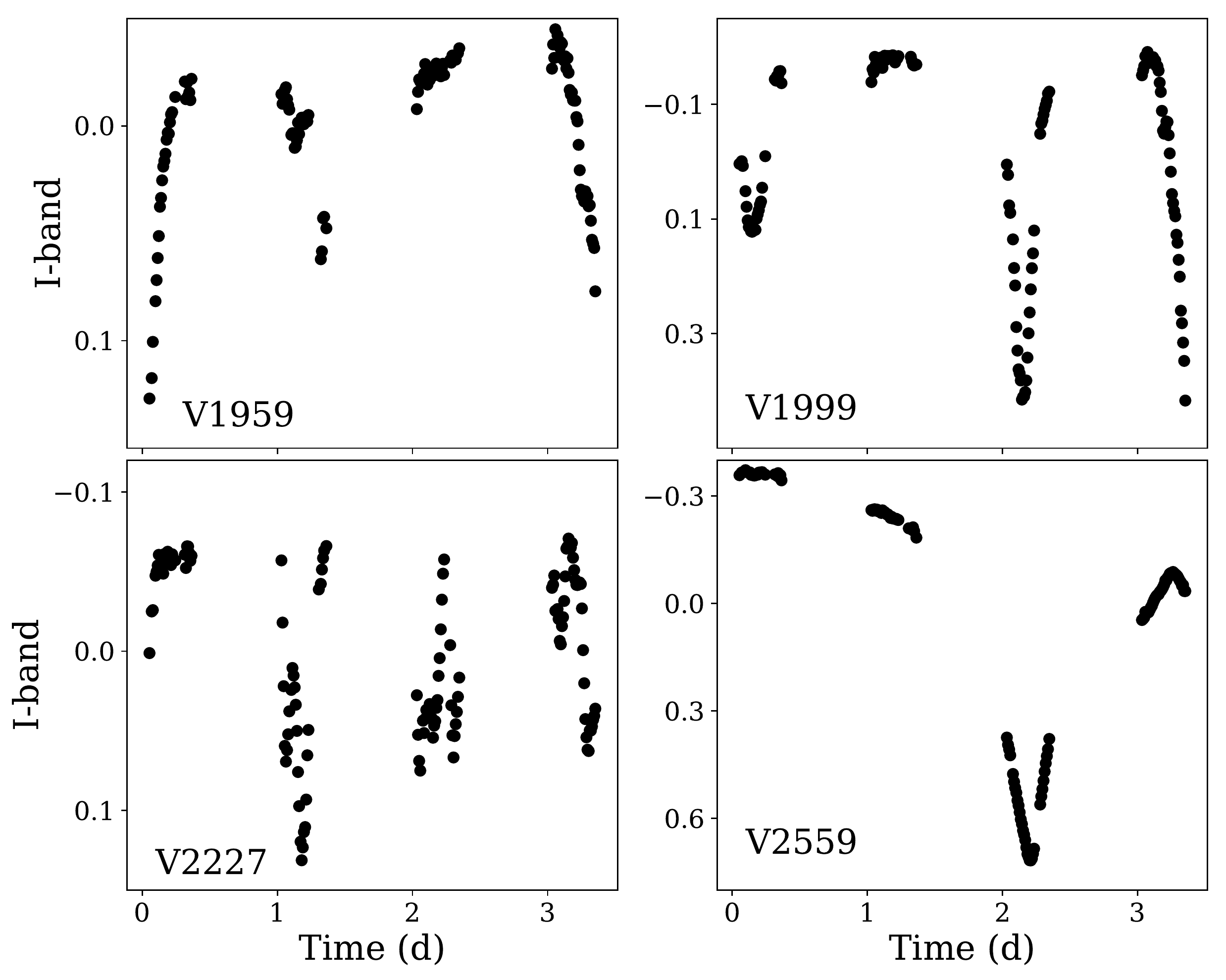}
\caption{Lightcurves for the four targets, as published in \citet{2005A&A...429.1007S}. The time frame covers the period 18-22 Dec 2001, with a null epoch of JD 2452262.0.}
\label{lc_se05}
\end{figure*}

The location of these four stars falls into the area of the cluster Collinder 70 that has been known since the 1930s. This cluster (which may not be gravitationally bound) is sparse and potentially overlapping with neighbouring populations \citep{2008A&A...485..931C,2017A&A...598A.124K}. The cluster catalogue by \citet{2005A&A...438.1163K,2009A&A...504..681K} lists a distance of 391\,pc, age of 5\,Myr, and negligble extinction, comparable to the cluster around $\sigma$\,Orionis slightly further south. The young stars in the Orion belt region belong to the larger Orion ecosystem, containing multiple overlapping populations with distances between 350 and 450\,pc and mean ages around 5\,Myr \citep{2018arXiv180504649K,2018arXiv180501008B}.

\begin{figure}
\includegraphics[width=8cm]{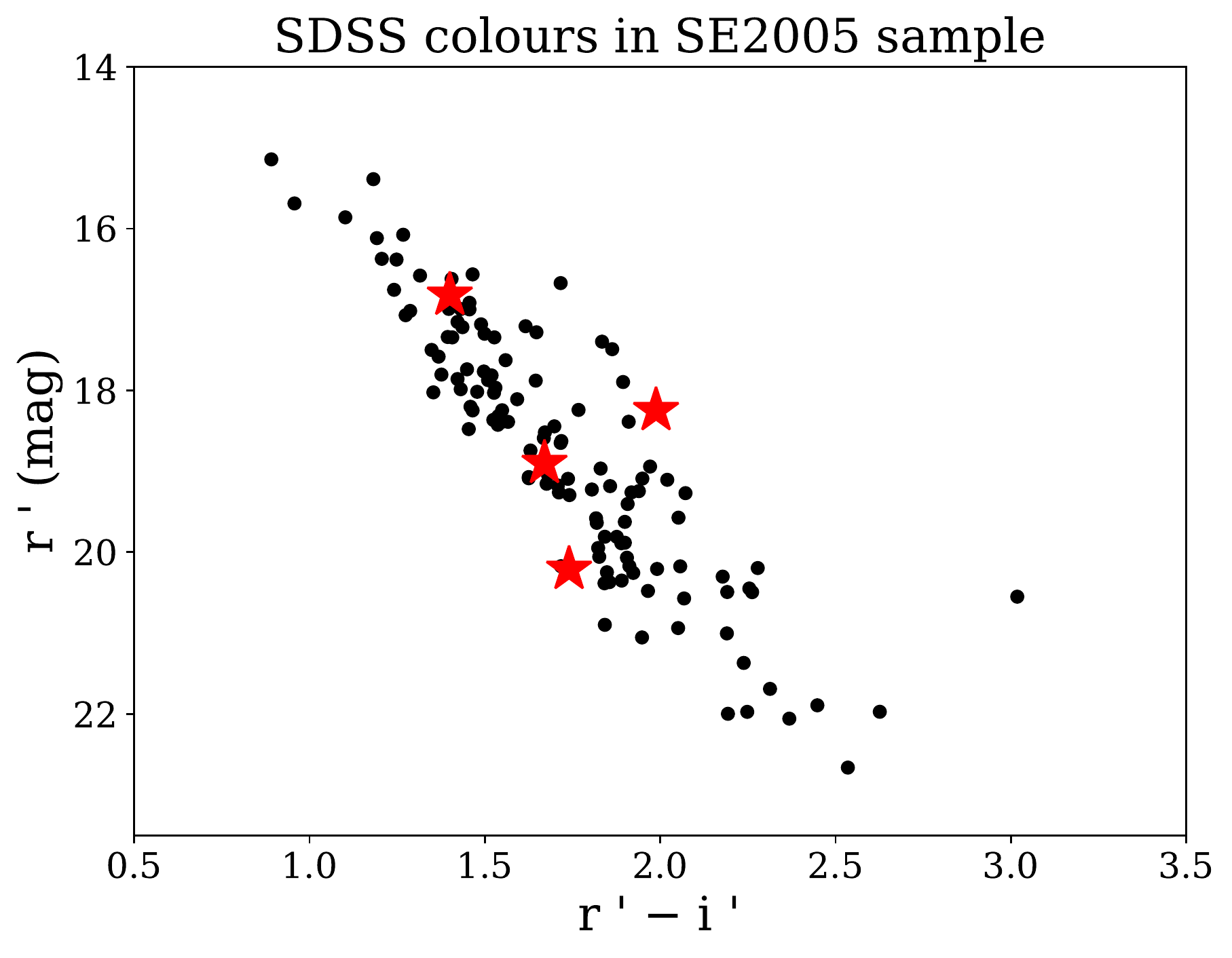}
\caption{SDSS DR12 colour-magnitude diagram for the full sample from SE2005. The four stars discussed in this paper are marked by big red stars. The star significantly redder than the bulk of the population is V1959.
\label{sdsscmd}}
\end{figure}

All four targets have M3-4 spectral types, as shown in a recent paper \citep{2016MNRAS.458.3118B}, corresponding to effective temperatures between 3400 and 3600\,K \citep{2014ApJ...785..159M}. This translates into masses between 0.3 and 0.5$\,M_{\odot}$ \citep{2015A&A...577A..42B}. In Fig. \ref{sdsscmd} we show the (r,r-i) colour-magnitude diagram for the four stars and the entire sample of very low mass objects published by SE2005, using data from SDSS DR12 \citep{2015ApJS..219...12A}. The Sloan magnitudes have a typical epoch of 2004.7. V2227 and V2559 fall within the bulk of the datapoints. V1999 appears to be slightly bluer or underluminous, whereas V1959 is redder or overluminous, deviations that may be caused by the variability.  

All four objects have an entry in the Gaia Data Release 2 \citep{2016A&A...595A...2G,2018arXiv180409365G}. Three of them are listed with parallaxes of 2.7\,mas, with errors between 0.2 and 0.3\,mas. This implies distances around 370\,pc, with large errorbars. The exception is V1999, with a (very uncertain) parallax of $1.3\pm 0.5$\,mas, which raises doubts about its association with the clustering in Orion. 
Coordinates and some stellar properties of the four targets are summarised in Table 1, together with their official name (Vxxx) and the identifier used in SE2005.

\begin{table*}
\caption{Target properties. The last column is the id number in SE2005.}
\label{table1}
\begin{tabular}{lllllllll}
\hline
Vnum & RA (J2000) & DEC (J2000) & $r$ (mag) & SpT & Teff (K) & $M$(M$_{\odot}$) & H$\alpha$EW (\AA) & SE2005\\
\hline
V1959 Ori & 05 34 27.011 & -00 54 22.75 & 18.25 & M2.9 & 3620 & 0.45 & 12 & 51\\ 
V1999 Ori & 05 34 39.477 & -00 54 32.14 & 20.21 & M3.9 & 3450 & 0.35 & 14 & 63\\ 
V2227 Ori & 05 35 12.791 & -00 36 48.54 & 18.90 & M3.3 & 3550 & 0.40 & 6 & 87\\  
V2559 Ori & 05 35 38.509 & -00 51 11.47 & 16.82 & M3.2 & 3570 & 0.40 & 3 & 120\\ 
\hline
\end{tabular}
\end{table*}

All four show H$\alpha$ in emission \citep{2016MNRAS.458.3118B}, but the EWs (3-14~\AA) are lower than what we would expect for ongoing gas accretion from a disk \citep{2003AJ....126.2997B}. The EWs in H$\alpha$ are stable over four observing nights within the uncertainties, another sign that the emission is caused by the stellar chromospheres, not by accretion. Photometric lightcurves in 2003 confirmed the high-amplitude photometric variations and the general lightcurve morphology observed in 2001 \citep{2016MNRAS.458.3118B}. 
V1959 and V2559 show evidence for excess emission at 8, 12, and 24$\,\mu m$, indicating the presence of circumstellar dust \citep{2016MNRAS.458.3118B}. However, the excess is very weak compared to the level seen in typical young stellar objects with accretion disks (Class II objects). For V2227 and V1999 the excess emission, if present, is not detected. A more in depth discussion of the infrared SEDs will be presented in this paper (see Sect. \ref{sed}). 

\section{New observations}
\label{newobs}

\subsection{Optical lightcurves from LCO}

We observed the four stars using 1-m telescopes in the LCO network \citep{2013PASP..125.1031B}, from late November 2015 to late March 2016, as part of program STA2015A-002. The observations were set up to visit each star at least once every 48 hours (subject to weather conditions), with each visit comprising a single measurement in the Sloan r- and i-band filters, with exposure times of 300\,sec. The central wavelengths of these two filters are 6215 and 7545\AA, slightly shorter than the SDSS band definition \citep{1996AJ....111.1748F}. The observing strategy allowed us to monitor the optical variability over timescales of days to months and obtain colour information. The images in the two filters are taken one after the other, i.e. 'quasi-simultaneously', with an epoch difference of about 5\,min.

The lightcurves were produced by aperture photometry on the LCO-pipeline reduced images with subsequent calibration. For the latter, we calculated a reference lightcurve from a set of 5-10 reference stars in the same fields, and subtracted this reference lightcurve from the target lightcurve. This corrects for effects of varying extinction and airmass. Second, from the same set of reference stars we calculate the offset between measured magnitudes and SDSS magnitudes. This offset depends slightly on colour $r-i$. Therefore we determine a linear fit between offset and colour and apply this fit to the target lightcurve. This shifts the lightcurve to the SDSS photometric system. The typical shift between calibrated and measured magnitudes is $\Delta m = 4.6$\, in the r-band and $\Delta m=4.3$\, in the i-band.

\begin{figure*}
\includegraphics[width=7.0cm]{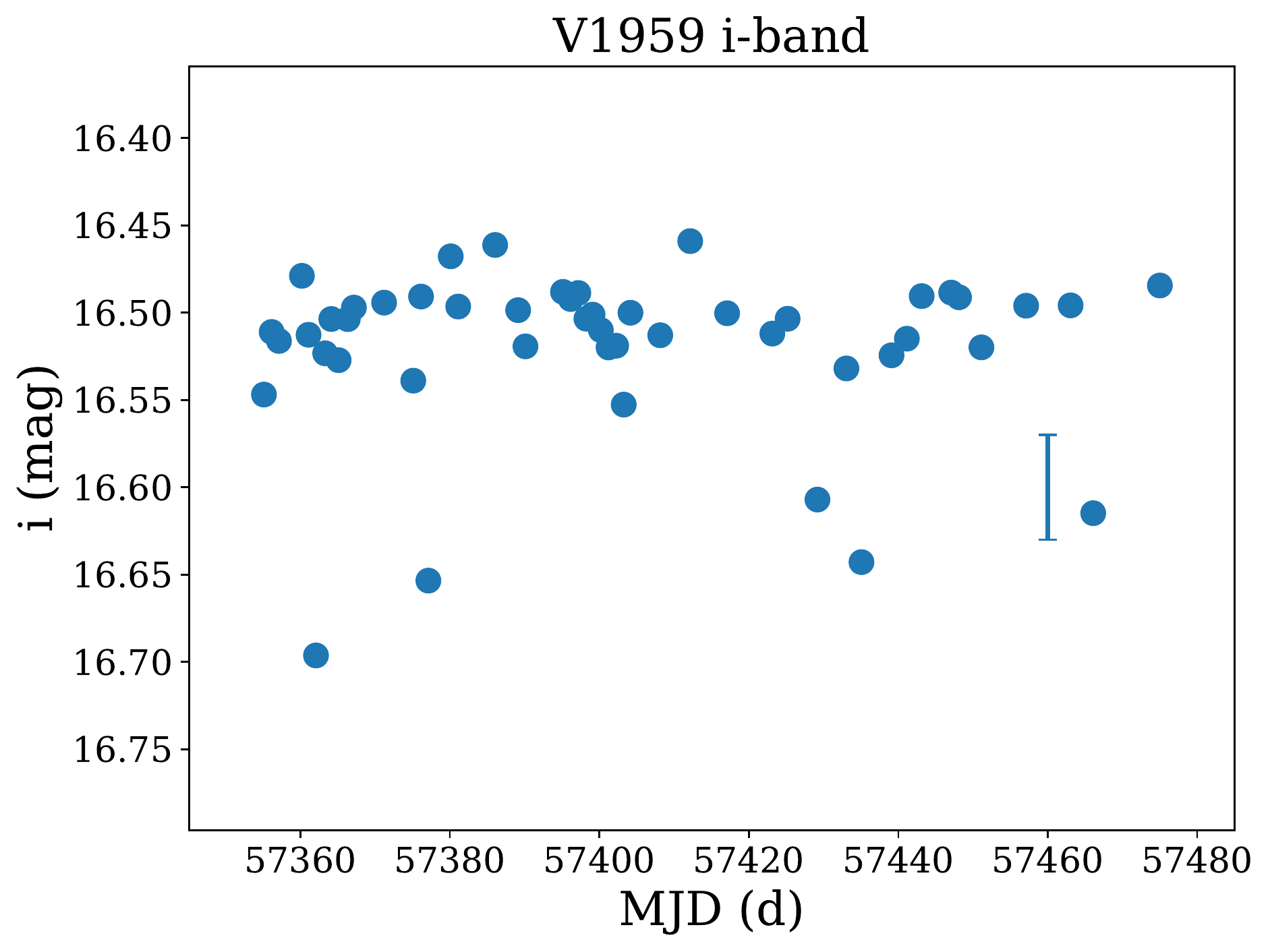}
\includegraphics[width=7.0cm]{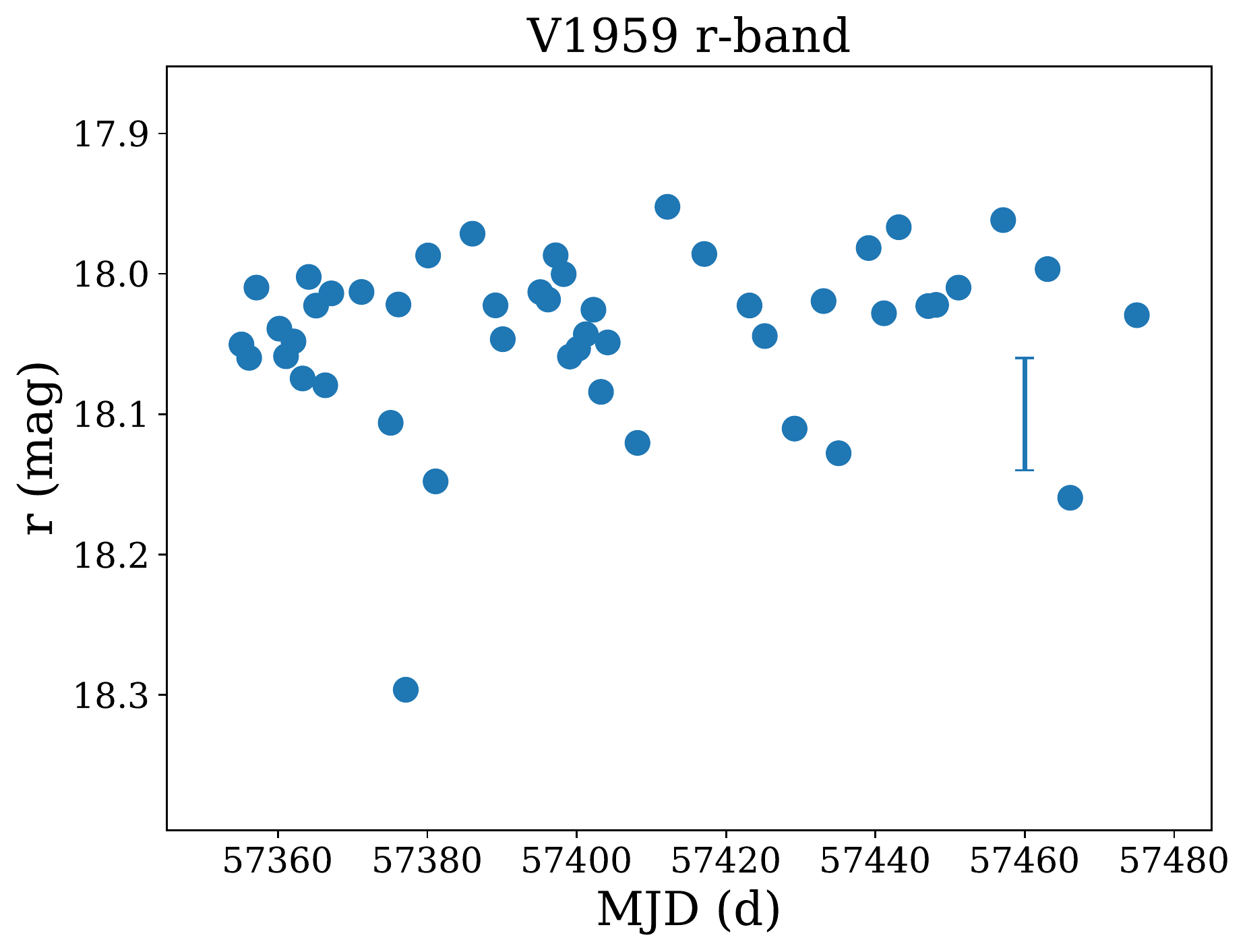} \\
\includegraphics[width=7.0cm]{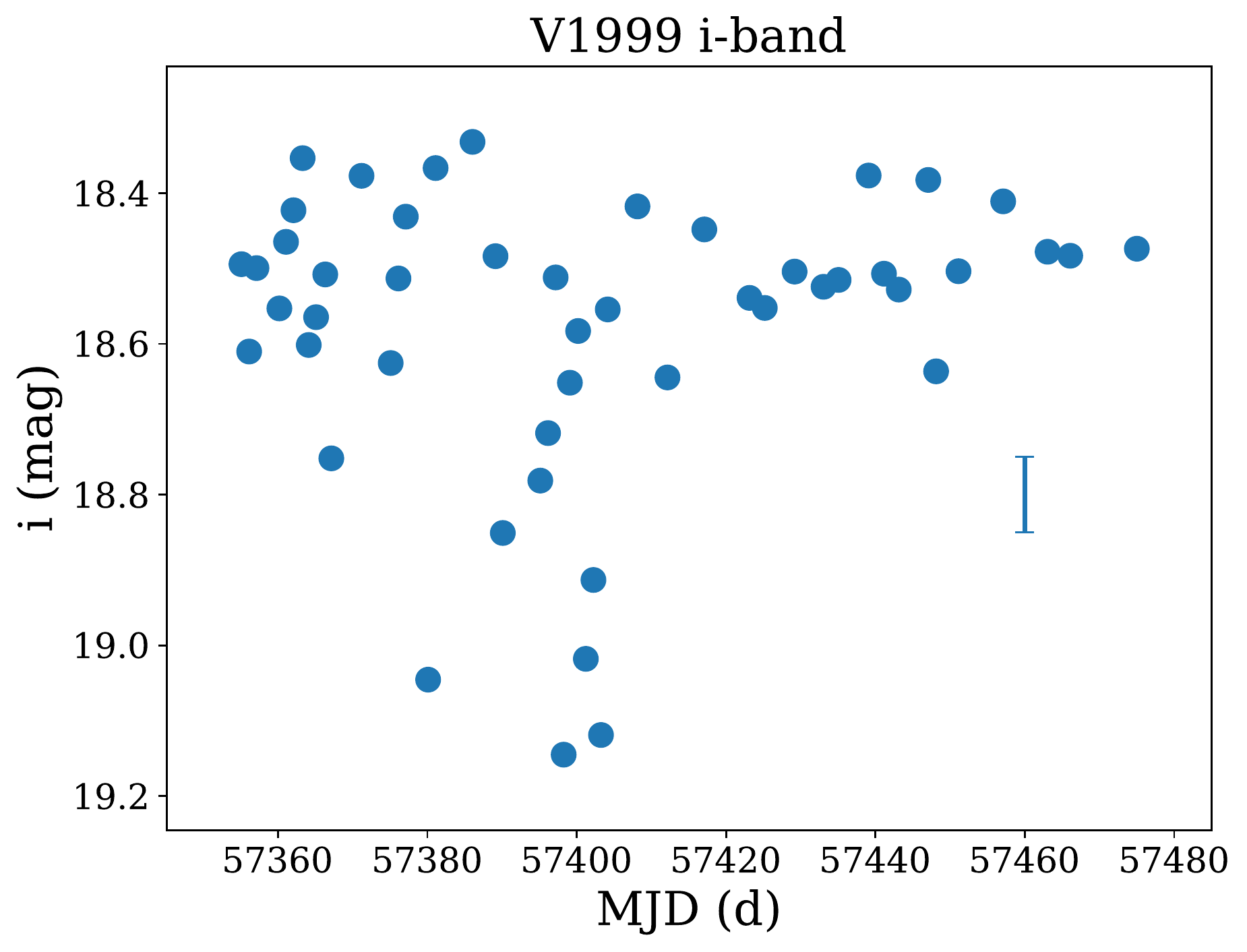}
\includegraphics[width=7.0cm]{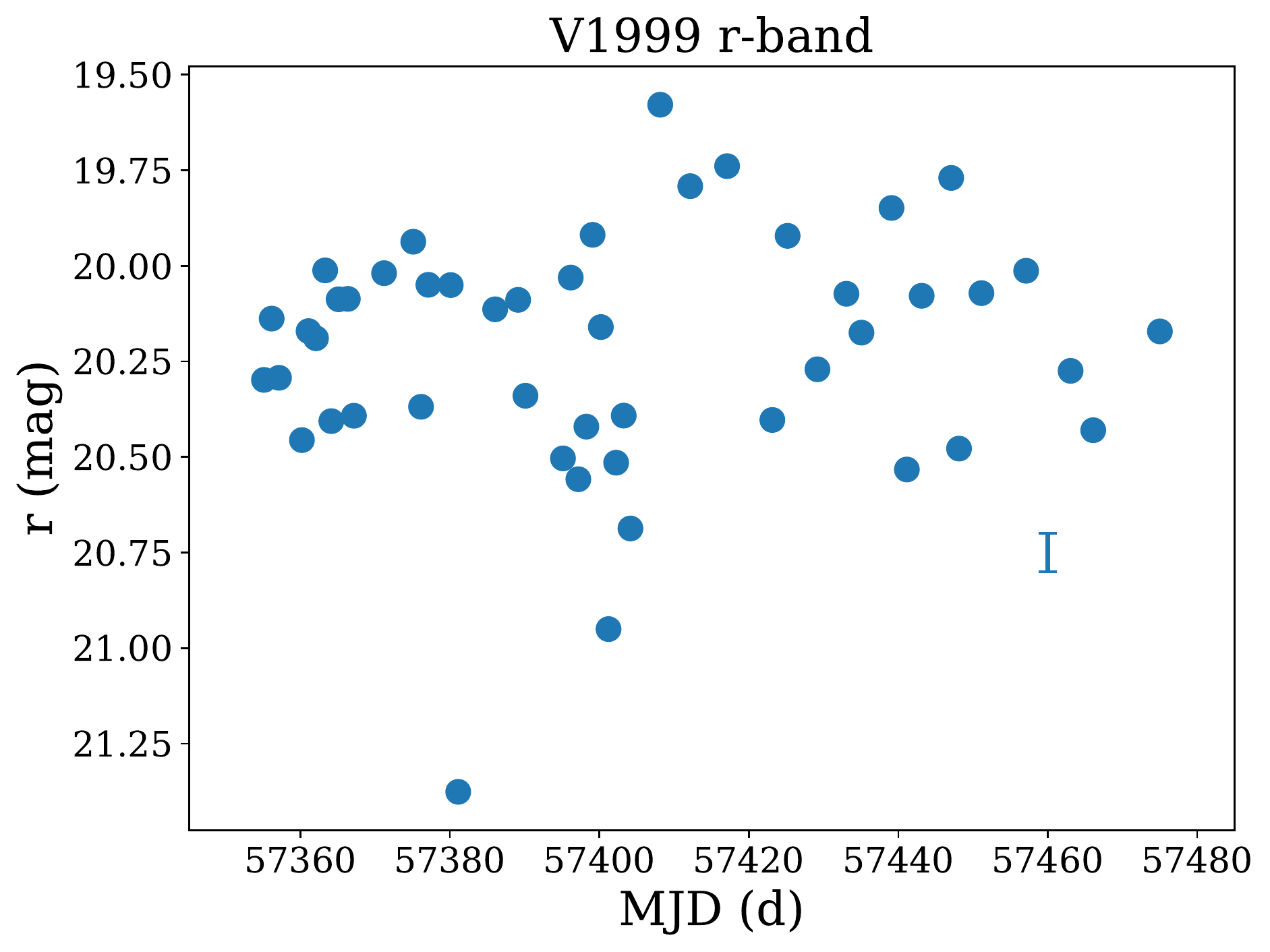} \\
\includegraphics[width=7.0cm]{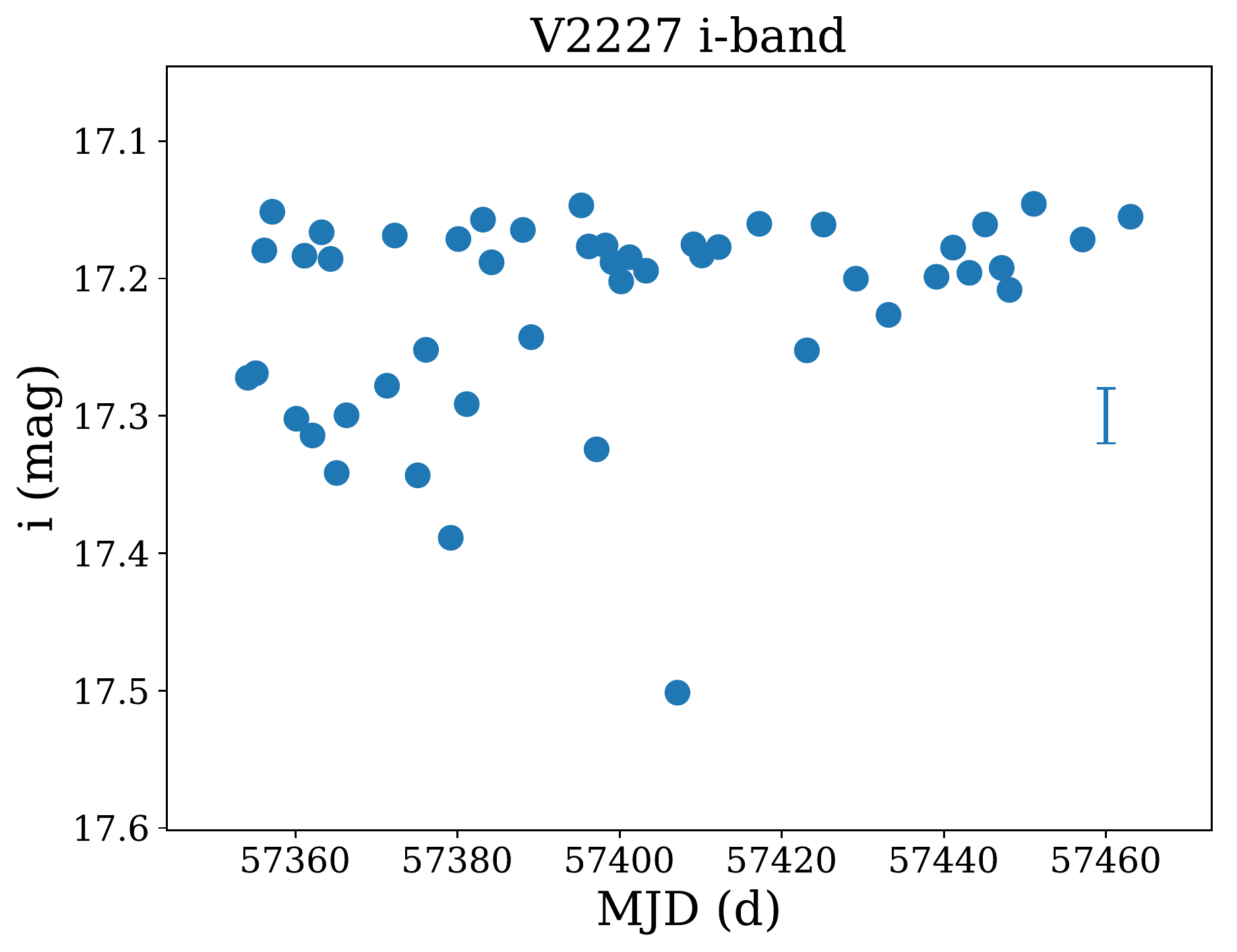}
\includegraphics[width=7.0cm]{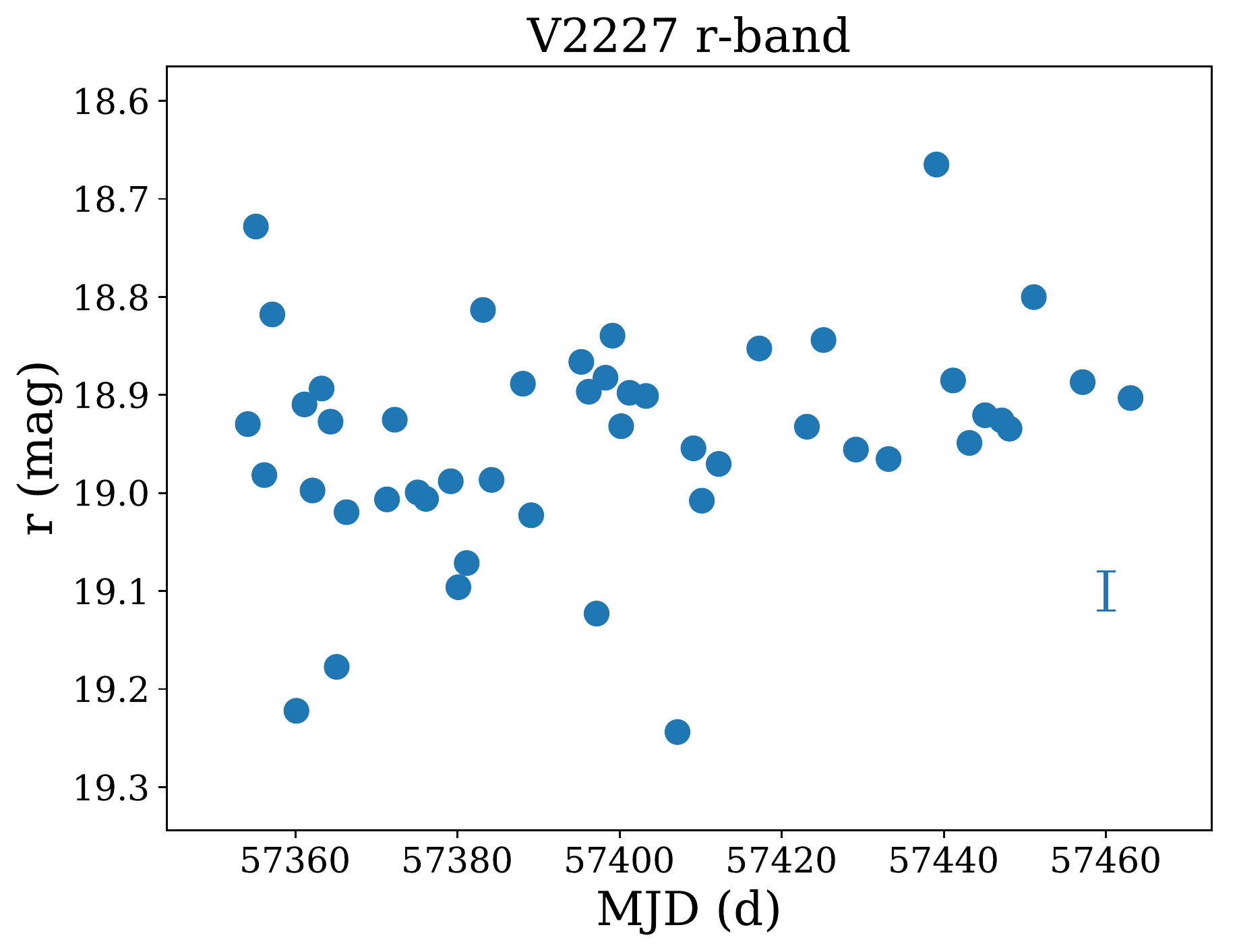} \\
\includegraphics[width=7.0cm]{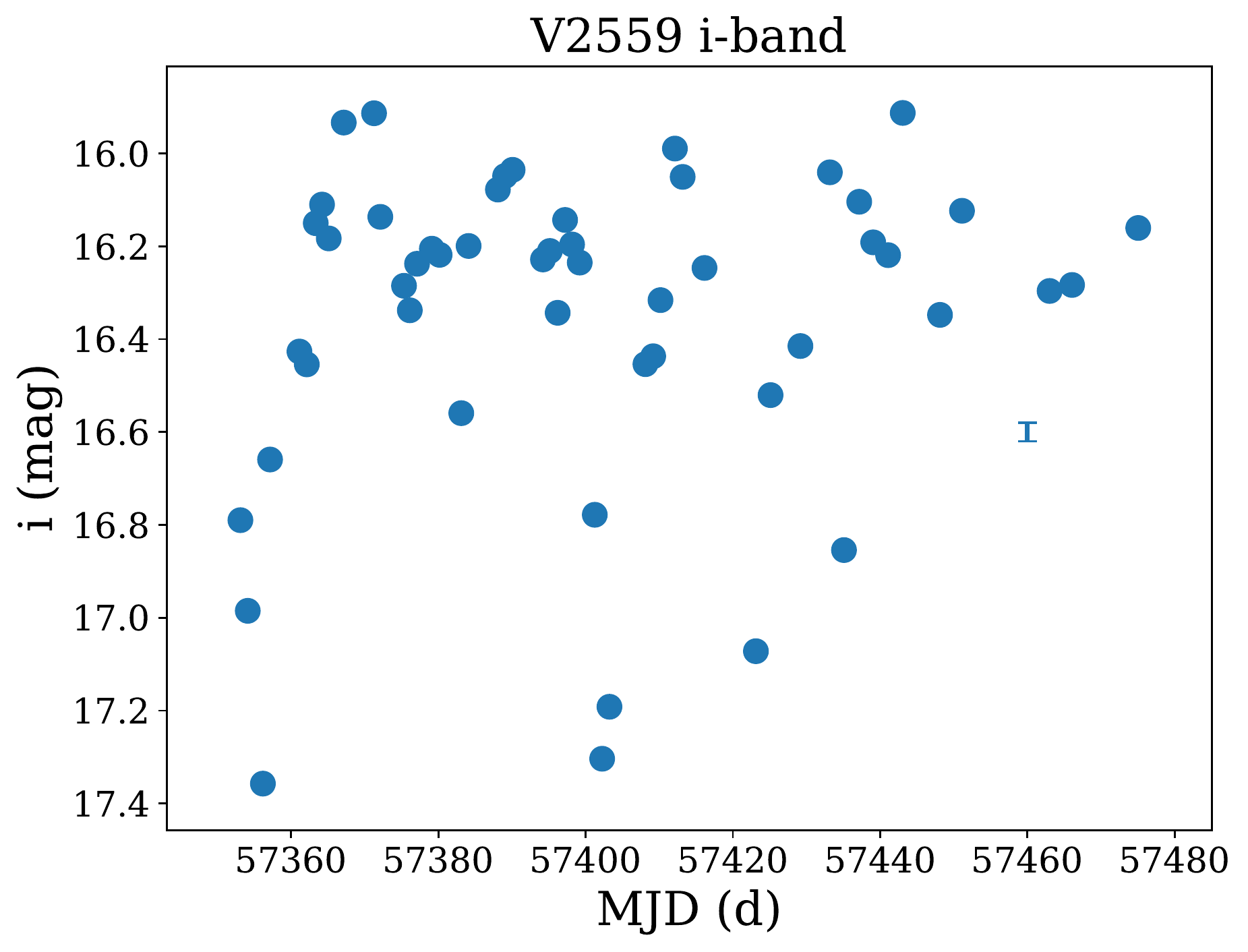}
\includegraphics[width=7.0cm]{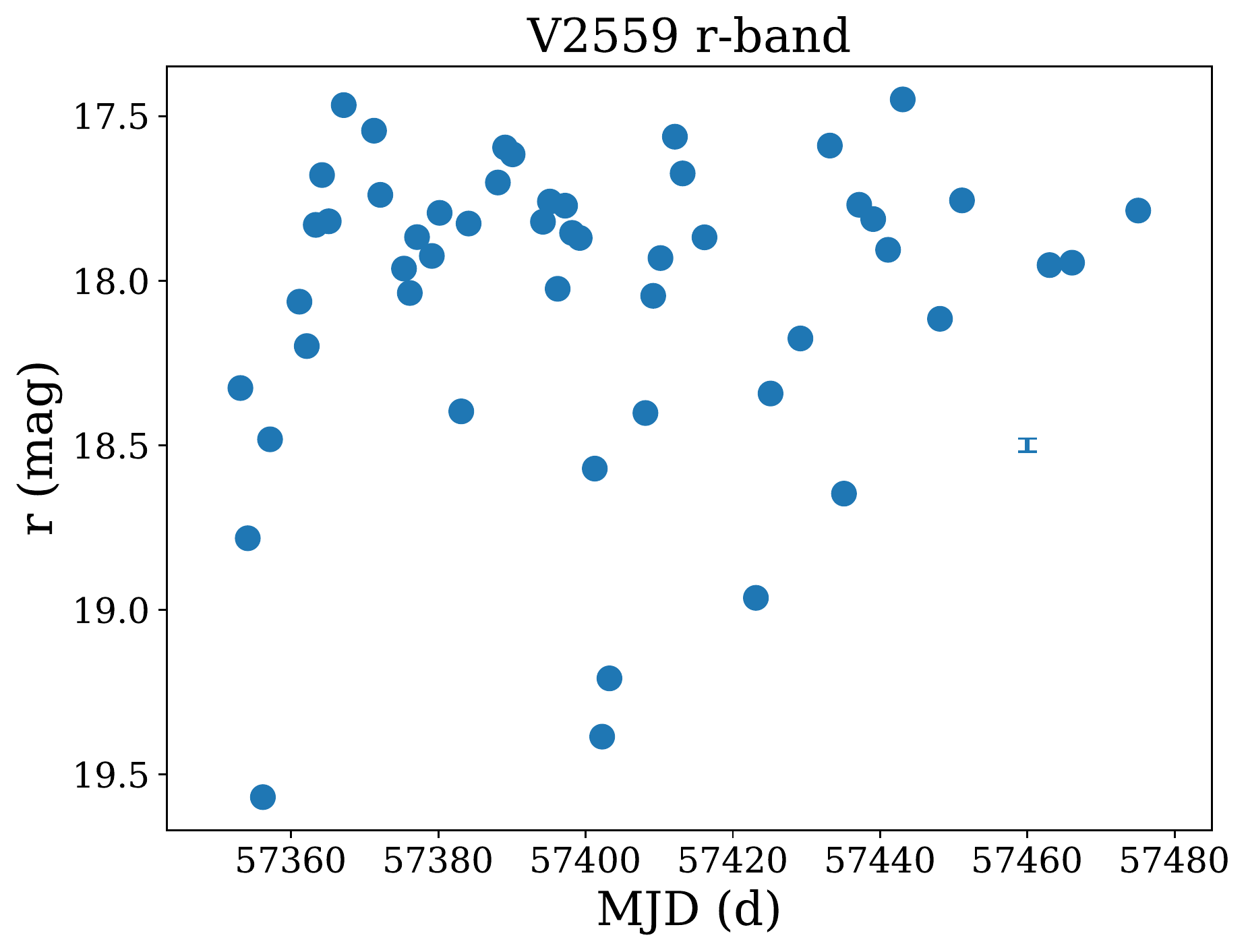} \\
\caption{Lightcurves from the Las Cumbres Observatory in the i- and r-bands for all 4 targets. Plotted is also a typical errorbar.}
\label{lc_lco}
\end{figure*}

Errors for the photometry are 2-5\%, estimated based on the standard deviation in lightcurves of reference stars with similar brightness. Given that targets are redder than the reference stars, there might be an additional calibration error due to non-linear colour terms. In Table 2 we summarise the lightcurve characteristics in r- and i-band for each object and filter. The lightcurves are shown in Fig. \ref{lc_lco}. 

Comparing the median magnitudes from the LCO lightcurves with the Sloan DR12 values (taken 10 years earlier) shows mostly good agreement. V2559 is brighter in the Sloan archive, by about one magnitude, indicating long-term changes in the flux level. V1959 is bluer in our photometry compared with Sloan and more in line with the bulk of the datapoints in Fig. \ref{sdsscmd}.

All four targets are found to be variable with amplitudes significantly exceeding the typical photometric error. The amplitudes increase with decreasing wavelength, for all four targets, as expected for reddening due to extinction. The lightcurves exhibit a clear cumulation of datapoints near the maximum, as expected for eclipses, although the sampling is too sparse to trace the individual eclipses. Compared to the SE2005 I-band datasets, covering only a few days, the amplitudes in the i-band are larger in all four cases. This could mean that the full scale of the variability is only apparent over long time windows. The LCO lightcurves confirm that the variability reported by SE2005 is not a transient phenomenon and is observable over more than a decade. 

\begin{table*}
\caption{Characteristics of the LCO lightcurves. $\Delta$ stands for the photometric amplitude, calculated as difference between minimum and maximum. $\sigma$ is the  standard deviation, $M$ the symmetry metrics (see Sect. \ref{lc_ana}), $N$ the number of datapoints.}
\label{table2}
\begin{tabular}{|l|lllll|lllll|}
\hline
object    & \multicolumn{5}{c|}{$r$-band}                & \multicolumn{5}{c|}{$i$-band}\\
          & median  & $\Delta$ & $\sigma$ & M  & $N$	 & median & $\Delta$ & $\sigma$ & M &  $N$   \\
\hline
V1959 Ori & 18.02   &  0.34    & 0.06 & 0.82  & 48	 & 16.50  &  0.23    & 0.05 & 1.17 	& 48	\\  
V1999 Ori & 20.17   &  1.80    & 0.31 & 0.43  & 48	 & 18.51  &  0.81    & 0.19 & 1.05	& 48	\\  
V2227 Ori & 18.93   &  0.59    & 0.11 & 0.37  & 48	 & 17.19  &  0.36    & 0.07 & 1.14  & 48	\\  
V2559 Ori & 17.87   &  2.12    & 0.47 & 1.02  & 51	 & 16.24  &  1.44    & 0.35 & 0.97	& 51	\\  
\hline
\end{tabular}
\end{table*}

\subsection{Optical lightcurves from CRTS}

\begin{figure*}
\includegraphics[width=16cm]{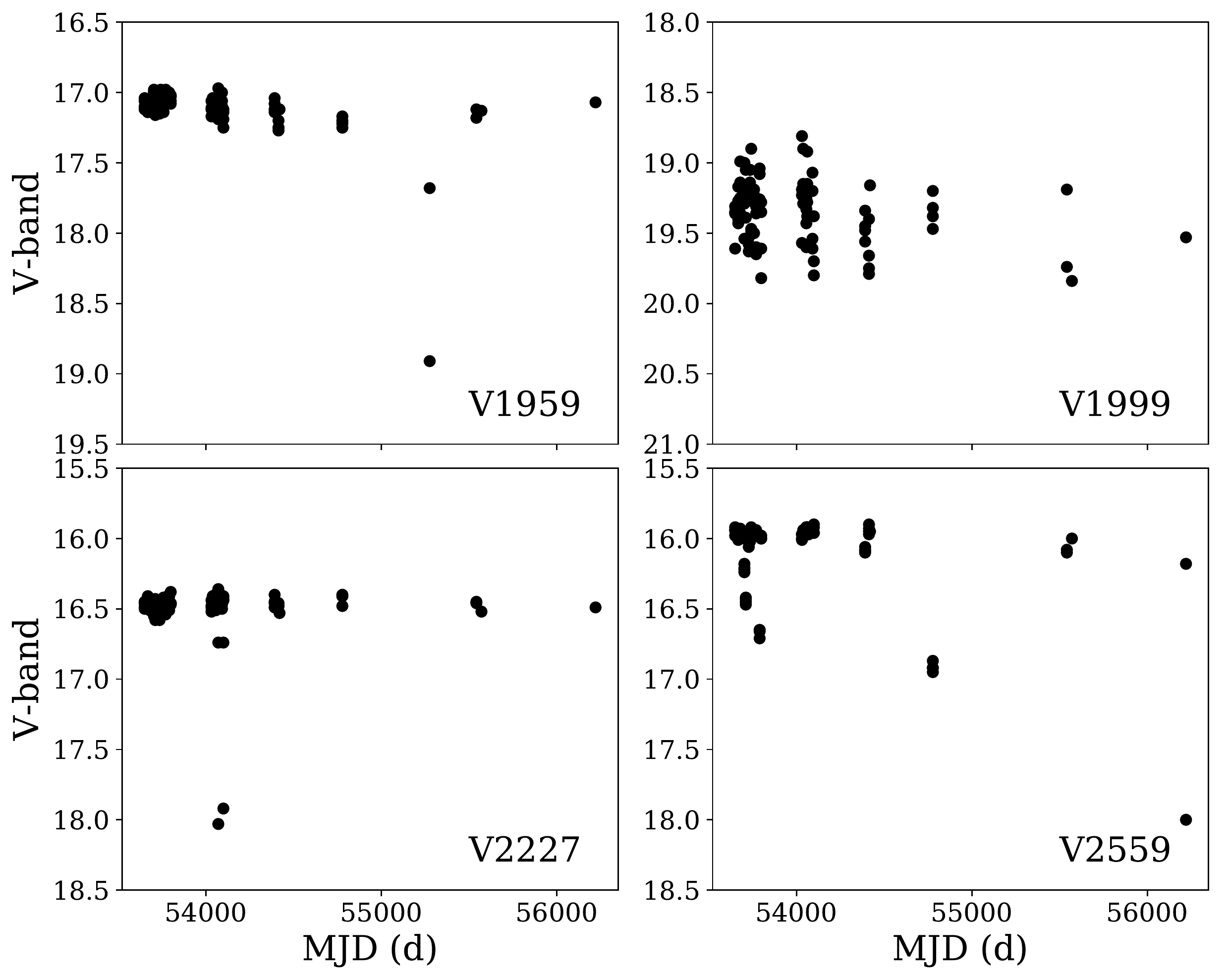}
\caption{Lightcurves for the four targets from the Catalina Real-time Transients Survey CRTS. The typical photometric error is 0.05-0.1\,mag, except for the V1999 data which comes with an error of 0.1-0.3\,mag.}
\label{lc_crts}
\end{figure*}

We downloaded the lightcurves from the Catalina Real-Time Transient Survey \citep{2009ApJ...696..870D} for all four stars. The CRTS photometry is in white light, but calibrated to a V-band. The lightcurves are reproduced in Figure \ref{lc_crts}. The survey contains about 90 epochs for each of our targets, between MJD 53000 and 57000, corresponding to years 2004 to 2014, roughly filling the gap between the published lightcurves and the new LCO data, albeit with limited cadence and limited
photometric precision. 

For V1959, V1999 and V2227 the error in the CRTS lightcurve is comparable or exceeds the standard deviation of the variability observed with other telescopes, therefore we cannot ascertain the long-term nature of these changes from the CRTS data. However, in V1959 and V2227, the CRTS lightcurves show some datapoints about 1.5-2\,mag fainter than the average lightcurve, which may point to occasional dramatic drops in the brightness, deeper than observed by SE2005 or in our LCO dataset. For V2559 the CRTS data shows a number of datapoints 0.5-2.0\,mag below the average, consistent with the other lightcurves. 

\subsection{Infrared lightcurves}
\label{ir}

We collected the infrared lightcurves from the WISE mission in the two bands centered at 3.6 and 4.5$\,\mu m$. We combined data from the 'AllWISE Multi-epoch Photometry Table' \citep{2010AJ....140.1868W} and from the 'NEOWISE-R Singe Exposure (L1b) Source Table' \citep{2011ApJ...731...53M}. Data for these objects was taken in 8 nights, between 2010 and 2016. V1999 and V2227 do not show significant signs of variations in these mid-infrared bands, but these are also the faintest objects with errorbars of 0.1\,mag or larger. For V1959 and V2559 there is some variability on timescales of months, but only marginally larger than the errors. We do not make use of the WISE lightcurves in the remainder of this paper.

\section{Lightcurve analysis}
\label{lc_ana}

In this section, we aim to gain insights into the physical nature of these systems from the properties of the lightcurves. 

\subsection{Symmetry}

As already pointed out in Sect. \ref{targets}, the optical data from SE2005 gives the impression that the lightcurves are asymmetric, in the sense that the objects spend most time near their brightness maximum. This would be indicative of eclipses as cause of the variability. To quantify the symmetry, we calculate the metric proposed in \citet{2014AJ....147...82C}:

\begin{equation}
M = (<d_{10\%}> - d_{med}) / \sigma
\end{equation}

Here, $<d_{10\%}>$ is the mean of the 10\% highest and 10\% lowest datapoints of the lightcurve, $d_{med}$ is the median of the full lightcurve, and $\sigma$ is the overall standard deviation. With this metric, values around 0.0 indicate symmetry, positive values are suggesting eclipsing objects, while negatives are flares or bursts. The SE2005 lightcurves for the 4 stars give $M$ in the range between 0.36 to 0.93, which confirms the visual impression. For the LCO i-band lightcurves, we obtain values between 0.4 and 1.2 (see Table \ref{table2}). Thus, based on the shape of the lightcurves, eclipses of some kind are a plausible explanation for the variability in all objects. This is confirmed by the appearance of the CRTS lightcurves (Figure \ref{lc_crts}). 

\subsection{Typical variability timescales}

Next we aim to quantify the typical timescales for the eclipses. A basic approach is to measure the epochs of minima and calculate the average separation between one minimum and the next. In principle this would give us a timescale for the gap between repeating events. Two lightcurves show well-sampled minima in the SE2005 data (see Fig. \ref{lc_se05}). For V2559, there is only one minimum, at epoch 2.2\,d, which is a lower limit for the separation between consecutive minima. V1999 has two fully sampled minima, separated by 2.0\,d. V2227 does not have minima that can be unambigously identified with the available sampling, but the average separation is significantly shorter than the coverage in one night ($<0.3$\,d). Finally, for V1959 no minimum is fully sampled, but three can be identified in the beginning of night 1, the end of night 2, and the end of night 4. Consequently, the typical separation is in the range of 1-2\,d. 

The lightcurves from 2003 \citep{2016MNRAS.458.3118B} also cover multiple nights, but have relatively few datapoints. Here, V2227 shows four short minima separated by roughly 1\,d. V1959 and V1999 do not have well-sampled minima in 2003, but the typical separation between maxima and minima is in the range of 1\,d. V2559 exhibits one minimum in the 2nd night, similar to the 2001 datasets. 
 
A period search on the 2001 lightcurves was conducted by SE2005, using Lomb-Scargle and CLEANed periodograms, in conjunction with a series of other tests. Three of them, V1959, V1999, V2559, show evidence for periodicity, with periods of 15.7, 15.5, and 82.5\,h, but the lightcurves, when plotted in phase to these periods, show substantial residuals, different from the more prevalent spotted stars with sinusoidal periods in their sample. For V2227 no period could be identified. We tried to identify periods in the LCO lightcurves as well, but found none. Since the sampling in the LCO data is sparse compared to the separation of eclipses, this might not seem surprising.

Altogether, this implies that the eclipses are not recurring in a periodic manner. Consequently, the eclipsing structure cannot be a singular homogenuous body, instead it has to be multiple bodies. 

\subsection{Eclipse depth}

The depth of the eclipses is variable as well, as demonstrated by our LCO lightcurves. In V1959 it ranges from 0.05 to 0.3\,mag (in the i-band) and perhaps up to 2\,mag, as seen in the CRTS lightcurve. For V1999 the typical depth is about 0.2-0.4\,mag, but a few datapoints in the LCO dataset indicate the existence of $>1$\,mag minima. For V2227 the range is 0.2\,mag up to (possibly) 1.5\,mag. V2559 has very deep eclipses, with a minima of 1\,mag in SE2005, and even deeper events in LCO and CRTS. Again this points to an eclipsing structure that is made of multiple components which are spatially separated. These structures have to cover a significant portion or maybe all of the stellar surface.

\subsection{Flat lightcurves}

The fraction of the time that is spent near maximum or near minimum gives information about the configuration of the eclipsing bodies in their orbit. In the SE2005 lightcurves, only V1999 and V2559 have a few hours of flat lightcurve near the maximum, in the 2nd and 1st night of the run, respectively. In the 2003 lightcurves published by \citet{2016MNRAS.458.3118B}, the only object with a discernible flat portion is V2227, with 70\% of the datapoints near maximum, which is the exception to the rule. On the longer timescales covered in the LCO dataset, a clear upper envelope of the lightcurve is visible in all stars. The fraction of datapoints significantly below this envelope is estimated to be 10\% in V1959, 20\% in V1999, 25\% in V2559, and 30\% in V2227 (taken from the i-band lightcurves). Taken together, this implies that the multiple eclipsing bodies are spread out azimuthally, giving rise to the notion that they are distributed in an extended structure, like a ring or a disk.

On the other hand, none of the objects has any significant flat phase near the minimum, i.e. after dimming they immediately start rebrightening again. This rules out that the eclipsing bodies are substantially smaller than the star, as in the case of exoplanets, or substantially larger, as in the case of AU-scale structures in protoplanetary disks (e.g., in RW Aur, see \citet{2016MNRAS.463.4459B}). Instead, the individual eclipses have to be caused by structures similar in size to the stars. A more precise evaluation of the sizes of the occulters requires more information about their nature and their orbital inclination. In particular, the upper limit of the structure that can be inferred from the lack of the flat bottom depends on the optical depth profile. Optically thin objects with a non-uniform density profile can cause eclipses without flat bottom while being larger than the stars.

\subsection{Colour changes}

\begin{figure*}
\includegraphics[width=8cm]{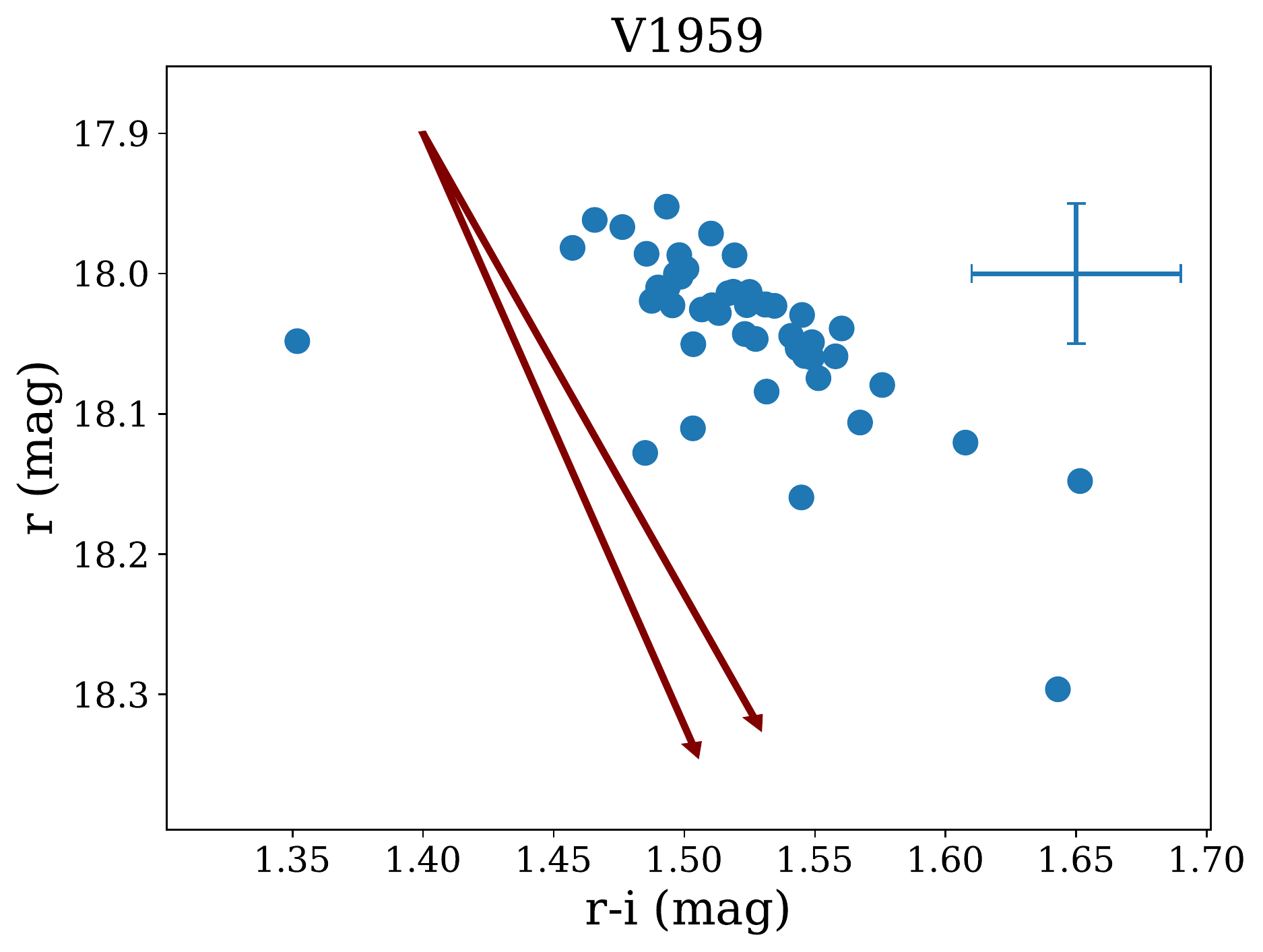}
\includegraphics[width=8cm]{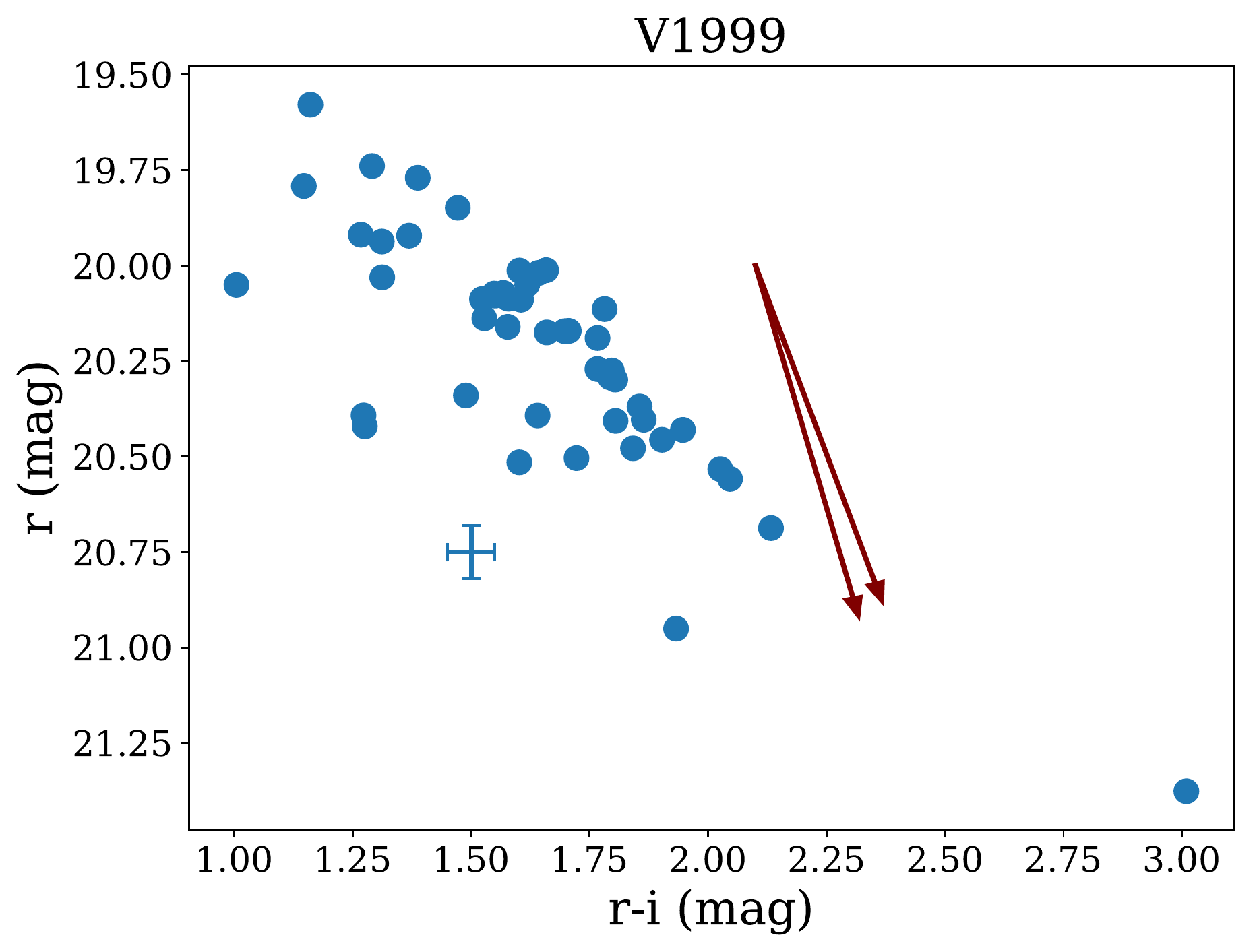}\\
\includegraphics[width=8cm]{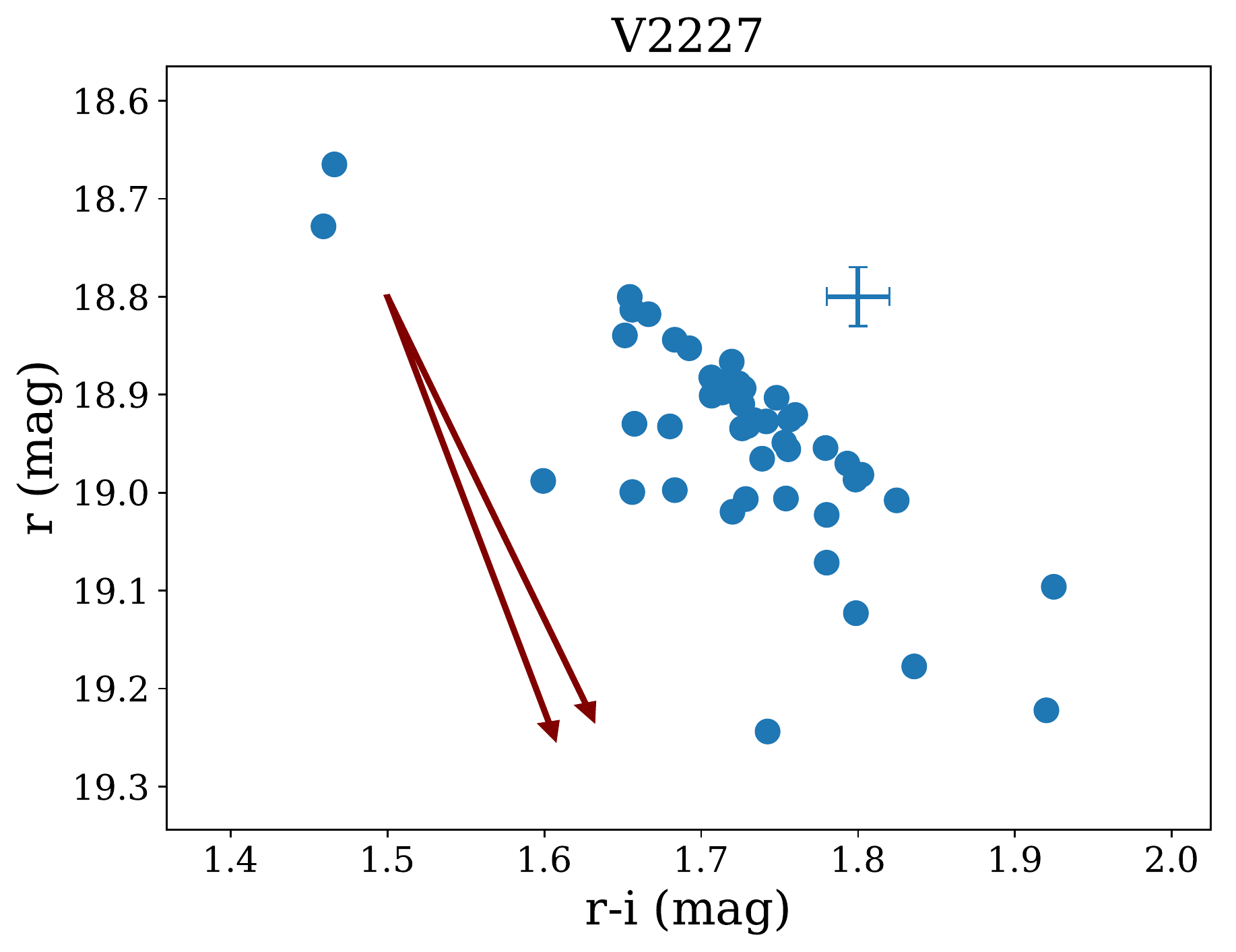}
\includegraphics[width=8cm]{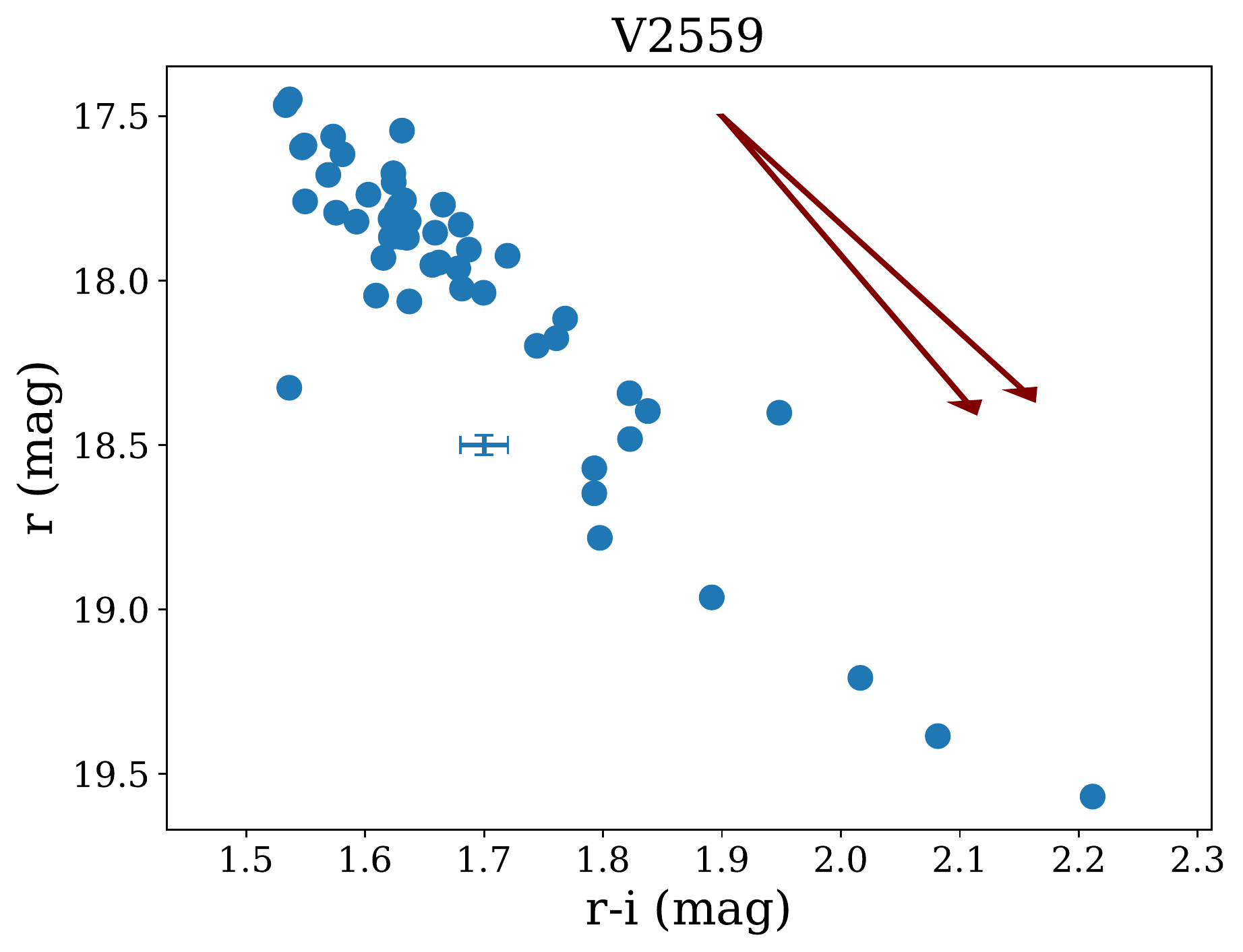}
\caption{Colour-magnitude diagrams produced from the LCO photometry. Overplotted are standard reddening vectors for $R_V=3.1$ and $R_V=2.0$, spanning $A_V=0.5$ in the plots in the left column and $A_V=1.0$ in the plots in the right column.}
\label{cmd_lco}
\end{figure*}

With the lightcurves from LCO we have for the first time broadband colour information on the brightness changes in the four stars. In Fig. \ref{cmd_lco} we show colour-magnitude diagrams derived from the LCO lightcurves. For these plots we make use of the fact that in most cases the LCO photometry provided quasi-simultaneous datapoints in $r$ and $i$, typically separated by 5\,min, a timescale short to most observed variations.

All four become significantly redder as they get fainter, by 0.2 to 2\,mag in $r-i$. This is the behaviour qualitatively expected for brightness changes due to variable extinction along the line of sight. For reddening to occur, the eclipsing structures have to be optically thin dust rather than optically thick planetesimals or asteroids. 

Note that this is at odds with the finding in \citet{2016MNRAS.458.3118B} -- in their spectral time series obtained in 2003 the spectra do not show evidence for reddening, with the exception of V2559. Although their spectral range (650 to 950\,nm) does not include the Sloan r-band, the colour changes shown in Fig. \ref{cmd_lco} would not have gone unnoticed. On the other hand, their time series extends only over 4 days, compared to the 4 months of the LCO dataset. As already pointed out in Sect. \ref{newobs}, the photometric amplitudes and also the associated colour changes increase with longer observational baselines, which could explain the discrepancy. It is also conceivable that the properties of the dust and hence the nature of the reddening have changed between 2003 and 2015. 
In Fig. \ref{cmd_lco} we show standard reddening vectors for interstellar dust, using $R_v=3.1$ and $R_v=2.0$ \citep{2002AJ....123..485S}. The reddening corresponds to changes in the line-of-sight extinction $A_V$ by 0.5 to 2\,mag. We note that the observed reddening is enhanced compared to the standard interstellar extinction law, in three out of four objects, the exception being V2559. This could be an indication that the reddening is caused by anomalous dust, although grain sizes larger than found in the ISM are expected to give the opposite trend (i.e. less reddening than in the ISM). As a cautious note when interpreting these diagrams, we re-iterate that the colour correction in the differential photometry might be incomplete, which could explain parts of the excess reddening (Sect. \ref{newobs}).

The change in the line-of-sight extinction provides constraints on the amount of occulting material. For an $A_V = 1$ and a dust absorption cross-section of $\kappa_{V} = 8.55 \cdot 10^3$\,cm$^2$g$^{-1}$ \citep{2003ARA&A..41..241D}, the column density is $N_{\mathrm{dust}} = \frac{1.086}{8.55\cdot 10^{-3}}$\,gcm$^{-2}$. Over the surface of a solar radius star, this gives a dust mass of $M_{\mathrm{dust}} \approx 10^{15}$\,kg. The precise value depends on the grain properties, the stellar sizes, and the total amount of extinction, which can only be approximated from our observations. Nevertheless, this exercise illustrates that only very little dust -- nine orders of magnitude less than an Earth mass -- is needed to cause the eclipses observed in these objects, so little that the eclipsing structures alone would be very difficult to detect in the infrared. The amount of dust is much smaller than what we expect to reside in the circumstellar disks around these objects (see Section \ref{sed}).

\subsection{Duration of ingress and egress}

The speed of the ingress and egress encodes information about the velocity and thus orbits of the eclipsing bodies. The SE2005 lightcurves with their dense sampling give the best constraints on ingress/egress timescales. Two of the stars, V1999 and V1959, go from maximum to minimum within one observing night (0.25\,d). V2559 takes $\sim 2$\,d for ingress or egress. V2227 has the fastest lightcurve changes with ingress/egress durations of about or less than 1\,h. This is consistent with the other datasets available for these objects. The fact that ingress and egress times are not changing significantly for a specific object, even when comparing lightcurves from different seasons, indicates that the eclipsing bodies have a similar velocity, and therefore a similar orbit. This would mean that they are not scattered throughout a disk, where a wide range of velocities is expected, but instead in a narrow ring of clumps having similar velocities.

Assuming an eclipsing body that is optically thick and has a sharp edge, a full ingress or egress requires the edge to travel the diameter of the star or the diameter of itself, whichever is shorter. We further assume that the structures responsible for individual eclipses are about the same size as the stars, as inferred above. At an age of $\sim 5$\,Myr, a M3-M4 star is expected to have a diameter of about $1\,R_{\odot}$. Under these assumptions, the eclipsing bodies have to move 200\,kms$^{-1}$ for V2227, 30\,kms$^{-1}$ for V1999 and V1959, and 4\,kms$^{-1}$ for V2559. For an optically thin occulter, as expected in these stars (see above), the velocity has to be higher. 

The velocities of the eclipsing bodies are related to their orbital separation via Kepler's laws, with $a = GM /v^2$. Assuming $M=0.4\,M_{\odot}$ for the stellar mass and neglecting the mass of the occulter, the orbital radii for V2227, V1959, V1999 and V2559 are 0.01, 0.4, 0.4 and 20\,AU. These numbers should be considered order of magnitude estimates. The estimate for V2227 is in the range of the dust sublimation radius, the shortest separation for dust grains to survive \citep{2002ApJ...579..694M,2006ApJ...645.1498S}. The eclipsing bodies in the star V2227 with ingress/egress times in the range of 1\,h should be very close to this limit. 

\begin{figure}
\includegraphics[width=8cm]{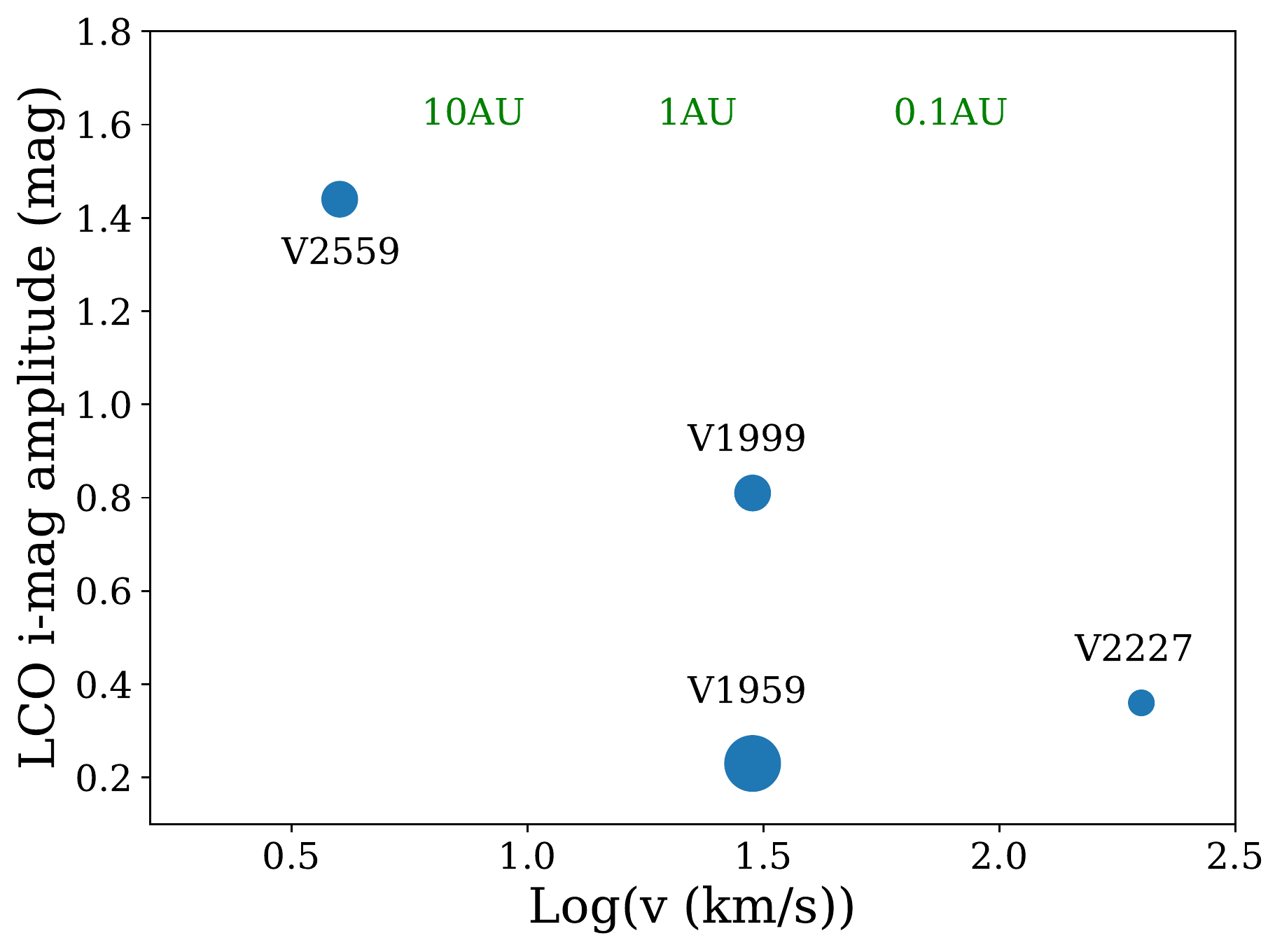}
\caption{Total i-band amplitude vs. orbital velocities for our four targets. Indicated at the top of the plot are corresponding orbital radii (assuming constant mass of $M=0.4\,M_{\odot}$). The symbol size indicates the type of IR excess: big for full/evolved disk, medium for evolved disk, and small for debris disk (see Sect. \ref{over}).}
\label{velo}
\end{figure}

In Fig. \ref{velo} we show the orbital velocities and indicative orbital radii as a function of i-band lightcurve amplitude, the latter taken from the long-term coverage of the LCO dataset (see Table \ref{table2}). The size of the symbols encodes information about the type of disk (see Sect.  \ref{over}). For V2227, V1999, and V2559 we see a clear trend, in the sense that increasing orbital separation is correlated with increasing photometric amplitude. The fourth object V1959 does not match that trend. Given the low number of objects, we postpone a deeper evaluation of this possible characteristic correlation until a larger sample is available.

\subsection{Secular evolution}

Overall, these four stars have now been monitored over about 15 years, from 2001 to 2016. For the first time we have an opportunity to comment on the long-term evolution of the observed eclipsing events. In all four stars, the variability is persistent over that timespan, indicating that the occulting structures are not transient in nature. If these structures are dusty clumps in orbit around the stars, their survival will be limited by shearing. Following \citet{2017RSOS....460652K}, the shear velocity across a clump of radius $R$ is $v_\mathrm{shear} = 3 R \Omega$. That means after one orbit the clump will be stretched by a factor of $6\pi$. The depth of the eclipses will be lower by the same factor. For the variability to persist, the eclipsing structures have to be either gravitationally bound or replenished from a dust reservoir. 

On the other hand, the similarity in the typical variability timescales and ingress/egress durations over different observing seasons points to stable orbits for the occulters, confirming the notion of a ring-like structure.

For V2227 there is good reason to believe that the frequency of the eclipses is changing over time. While the star shows continuous fluctuations in 2001, it is only seen in eclipse about once every night in 2003. The LCO lightcurves shows a number of dips in the first half, but few in the second half. It is plausible that the occulting structure is evolving over timescales of months to years. For this star, in addition to shearing any dust clump would be subject to sublimation by the heating from the star, as pointed out above. 

On the other hand, there is some evidence for stability for object V2559. As noted by \citet{2016MNRAS.458.3118B}, the morphology of the lightcurve of V2559 is remarkably similar in 2001 and 2003. Given the large orbital radius of 20\,AU inferred from the ingress/egress timescales (see above), these two events cannot be caused by the same occulter. This may indicate that different occulters in the ring have a similar structure. We have earlier mentioned the possible long-term changes in the brightness in this object (Sect. \ref{newobs}).

\subsection{Summary}

The four stars show irregular eclipses in optical light. The lack of a periodic pattern, the variable depth in the eclipses, and the fact that the stars are seen in eclipse for a large portion of the time points to a multitude of eclipsing bodies, spread out over a wide area, intercepting the emission from the star along the line of sight. Typical timescales in the eclipses are consistent for any individual object, i.e. the bodies all have similar orbits. Modest reddening equivalent to $A_V = 0.5-2$\,mag indicates that the occulters are optically thin, i.e dusty clouds, rather than solid bodies. The absence of a flat bottom in the eclipse means that the clouds have to be similar in size to the stars, or somewhat larger in case of a spatially varying optical depth. 

\begin{figure}
\includegraphics[width=8cm]{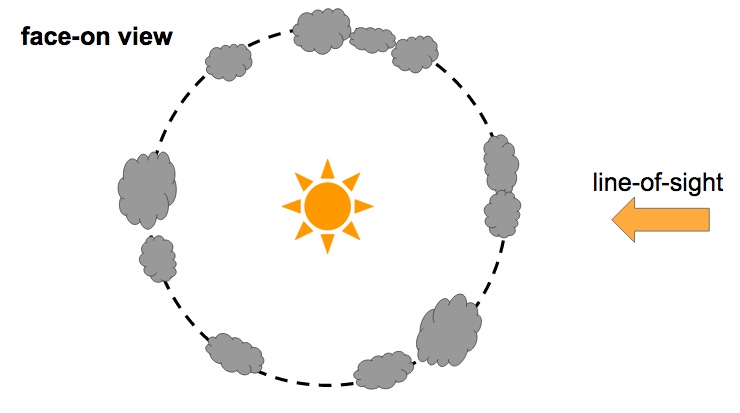}
\includegraphics[width=8cm]{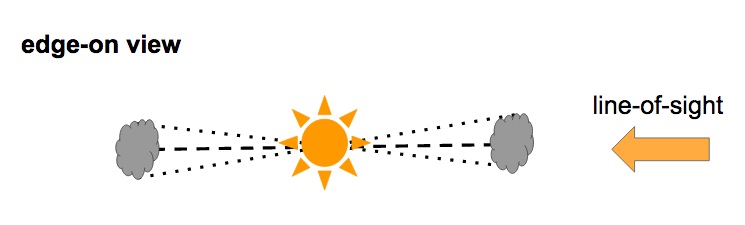}
\caption{Simplified sketch, not to scale, of the proposed geometry for the dusty rings, in two orientations. The dashed line indicates the plane of the orbit for the clouds in the rings. The dotted lines in the edge-on sketch indicates an opening angle of $\sim 20$ deg, see Sect. \ref{sed} for a motivation of that value.
\label{sketch}}
\end{figure}

Taken together, we explain the variability with the presence of a ring made of optically thin dust clouds, as illustrated in the sketch in Fig. \ref{sketch}. The radii of the rings are between 0.01 and 20\,AU in our sample. The dust mass of individual clouds is a small fraction of the mass of Earth and a negligible fraction of the dust mass typically found in circumstellar disks. The orbiting dust clouds are expected to be either gravitationally bound or transient features. In the latter case, the rings have to be replenished continuously.

\section{Spectral energy distributions}
\label{sed}

The lightcurves discussed in Sect. \ref{lc_ana} strongly suggest eclipses by circumstellar dust as cause of the variability. Therefore it makes sense to examine the infrared spectral energy distributions (SED). 

\subsection{Overview}
\label{over}

\begin{figure*}
\includegraphics[width=16cm]{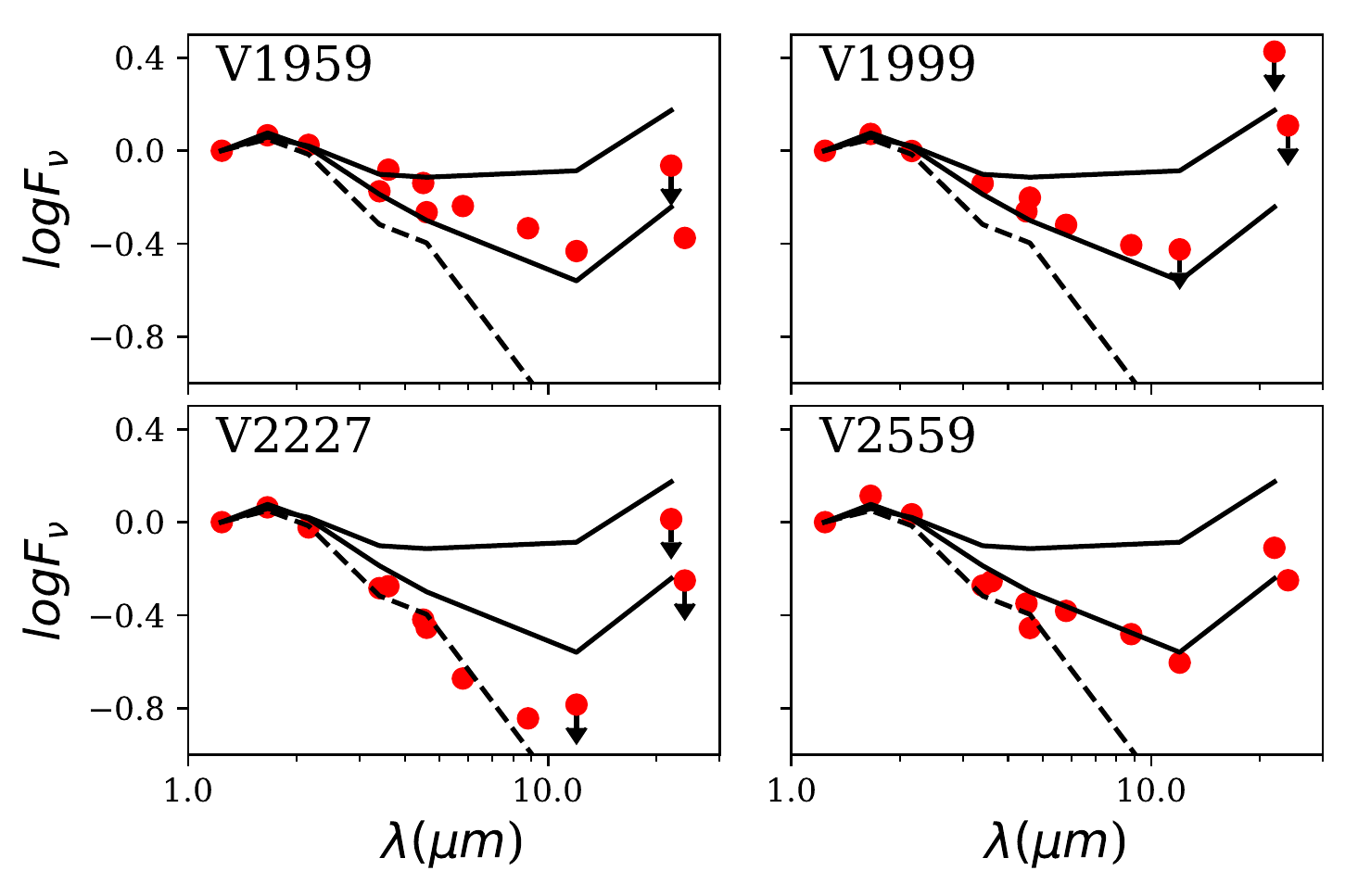}
\caption{Normalized spectral energy distributions $F_\nu / F_\nu(J)$ for the 4 stars in comparison with a templates (solid lines: UscoCTIO 33 and UscoCTIO 13, dashed line: AU\,Mic).
\label{sed1}}
\end{figure*}

All four objects have entries in the 2MASS \citep{2003yCat.2246....0C} and AllWISE databases \citep{2014yCat.2328....0C}, and have also been observed by Spitzer/IRAC and MIPS (as part of project \#50360, PI Briceno, \citet{2014MNRAS.444.1793D}). In Fig. \ref{sed1} we plot the available SED for these stars, normalised to the J-band flux at 1.25$\,\mu m$. While V2559 and V1959 have multiple flux measurements for $\lambda > 10\,\mu m$, the remaining two targets currently only have upper limits in this wavelength domain.

In Fig. \ref{sed1} we compare with the SEDs for three template objects at similar ages and spectral types. In solid lines, we show SEDs for two M3-M4 stars in Upper Scorpius, classified by \citet{2012ApJ...758...31L} as 'full' disk (UScoCTIO 33) and 'evolved' disk (UscoCTIO 13). The dashed line is the SED for AU\,Mic, a member of the $\beta$\,Pic moving group with early M spectral type. AU\,Mic hosts a debris disk seen close to edge-on  (e.g., \citet{2005ApJ...634.1372C}); as most debris disks, the SED is photospheric up to 24$\,\mu m$, but has excess at longer wavelengths \citep{2009ApJ...705.1646C}.

With the exception of V2227, the stars have clear excess emission in the mid-infrared. V1959 has an SED between full and evolved disks. V1999 and V2559 resemble the evolved disk. Finally, V2227 has an SED that is comparable to AU\,Mic's debris disk, perhaps the only difference being a small excess at 8$\,\mu m$. 

Thus, dust has to be present in close proximity to the stars, but none of them has a full primordial disk. We recall that these stars do not show evidence for accretion, i.e. the disks are presumably not gas-rich. The four disks in our sample all show a degree of evolution, with reduced emission at infrared wavelengths compared to primordial disks. Possible reasons for the evolution can be dust depletion and/or settling, perhaps in the case of V2559 combined with an inner opacity hole. The most extreme case is V2227 which might be best classified as debris disk, although its flux at 8$\,\mu m$ and the upper limits at longer wavelengths still leave plenty of room to hide circumstellar dust.

The comparison objects in Fig. \ref{sed1} have all been observed with high sensitivity in the submm-mm domain. From these observations we know that the dust mass in their disks ranges from 3.8$\,M_{\mathrm{Earth}}$ (UScoCTIO 33) \citep{2016ApJ...827..142B} to 0.01$\,M_{\mathrm{Earth}}$ (AU\,Mic) \citep{2015ApJ...811..100M}. Submm-mm observations for the $\epsilon$\,Ori stars would be challenging, due to their large distances, but based on their currently known SEDs, it seems reasonable to assume that they host disks with dust masses in the range of a fraction of an Earth mass as well, i.e. on the order of magnitude of $10^{24}$\,kg. This is by many orders of magnitude larger than the dust mass estimate for the bodies responsible for the eclipses seen in the lightcurves (see Sect. \ref{lc_ana}). Thus, the postulated rings causing the variability constitute only a minor fraction of the circumstellar dust in these sources.

\subsection{IR excess in the $\epsilon$\,Ori population}

\begin{figure}
\includegraphics[width=8cm]{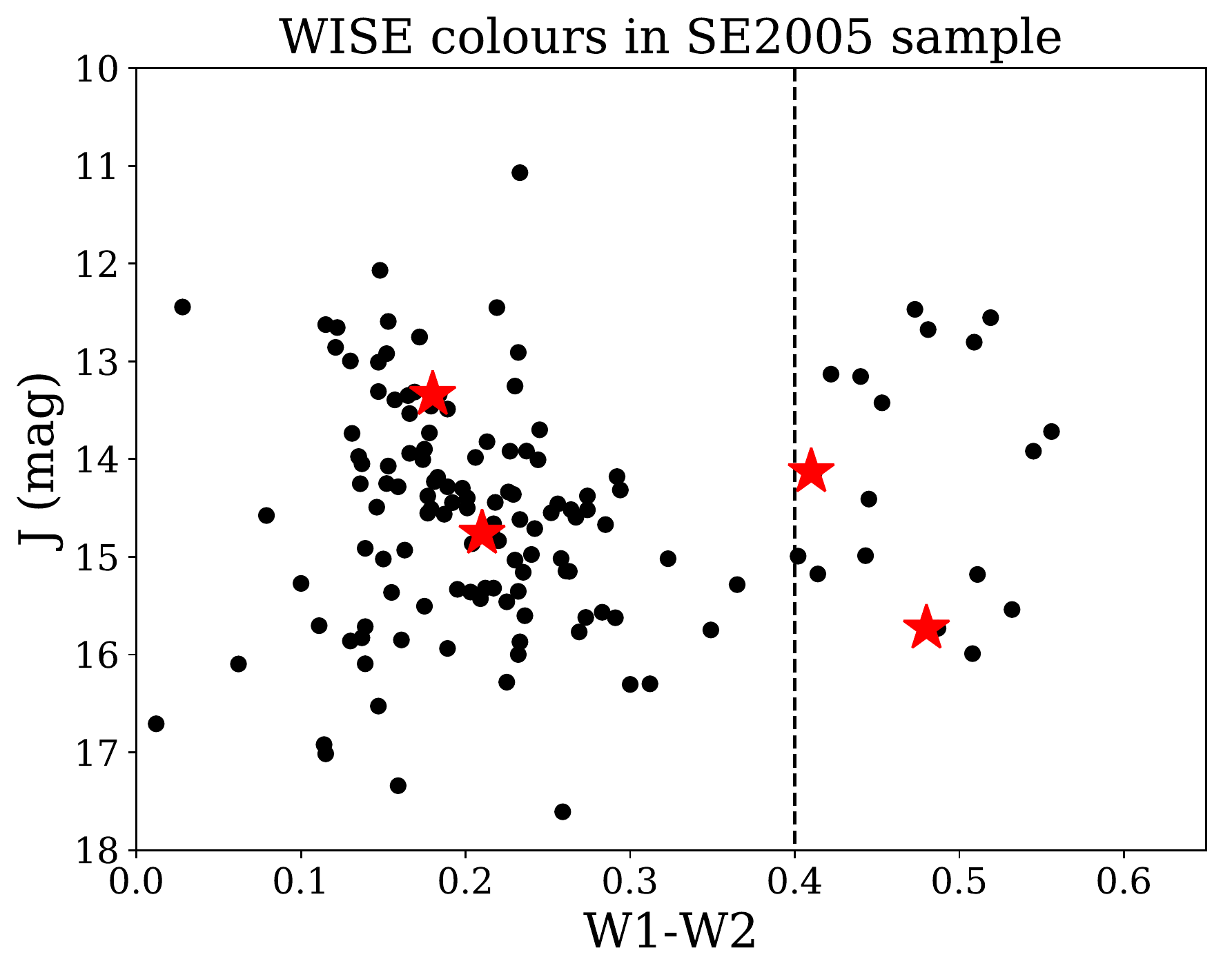}
\caption{WISE colour in the full sample from SE2005. The colour plotted in x is W1 (3.6$\,\mu m$)
minus W2 (4.5$\,\mu m$). On the y-axis, the 2MASS J-band magnitude is shown (1.2$\,\mu m$). The dashed line 
delineates the typical separation between objects with colour excess and without \citep{2013MNRAS.429..903D}. The four stars discussed in this paper are marked with red stars.
\label{sed3}}
\end{figure}

To put the four unusual stars in context, we estimate the frequency of low-mass stars with disks in the cluster around $\epsilon$\,Ori. We cross-checked the catalogue of objects published by SE2005 with the AllWISE database \citep{2014yCat.2328....0C}. Of the 143 objects, 137 have an unambiguous entry in WISE. Most of these only have upper limits or a very poor detection with large errorbars in channels W3 and W4, i.e. wavelengths of 12 and 22$\,\mu m$. Only 19 are solidly detected in W3 with errors $<20$\% and magnitudes $<11$, those are good candidates for objects with disks -- among them V1959 and V2559. The colours calculated from W1(3.6$\,\mu m$) and W2(4.5$\,\mu m$) show a clear bimodal distribution, with 18-20 objects being redder than what is normally expected for objects without disks (see Fig. \ref{sed3}). Among these are V1999 and V1959. 

We note that the SE2005 catalogue was selected from photometry alone, and may be affected by contamination from red stars in the background. SE2005 estimate a contamination rate of 16\%, by comparing with a Galaxy model. Some additional contamination by extragalactic unresolved red sources is possible. These contaminating objects are not expected to have mid-infrared excess. Accounting for contamination, the disk fraction among young objects in the SE2005 sample is $\sim 19 / (137 * 0.84) = 17\%$.\footnote{It is coincidental that this number is close to the estimated contamination rate.} Not all stars with disks at this age will necessarily have excess at 3-5$\,\mu m$, as evidenced by the SED of V2227 (see Fig. \ref{sed1}), i.e. the actual disk fraction including stars without colour excess at 3-5$\,\mu m$ is likely to be somewhat higher. For comparison, in the similarly aged Upper Scorpius association, the fraction of early M stars with either primordial or debris disks is in the range of 30\% \citep{2009ApJ...705.1646C,2012ApJ...758...31L}. For a population of young stars in Orion located southwest of $\epsilon$\,Ori, a disk fraction of 13\% has been found for early M dwarfs \citep{2007ApJ...671.1784H}. 

We showed in Section \ref{over} that at least three out of four of the highly variable stars have IR excess due to disks, although only two at 3-5$\,\mu m$. The fourth may also have excess emission at longer wavelengths. The WISE colour analysis for the full SE2005 sample demonstrates that our variable objects share this characteristic with a sizable population of similarly aged stars. Among objects with WISE IR excess at 3-5$\,\mu m$, the fraction of those with irregular eclipses is in the range of 12\% (2 of 19). Thus, we predict that large-scale time-domain surveys covering regions of similar age should find a substantial number of objects of this kind. The lightcurves produced by the Gaia mission are a promising resource to reveal larger numbers of these kind of variable stars, although they do not have the cadence to resolve the eclipses \citep{2018A&A...618A..30H}. We note that this type of variability would remain undetected in Gaia Alerts, simply because the objects are continuously variable.

The fraction of objects with irregular eclipses gives us additional information about the configuration of the occulting bodies. If the dust is confined to a disk-like structure and if inclinations are random, the opening angle of the disk is $\alpha = f \times 180$\,deg, where $f$ is the fraction of stars that shows evidence for variability due to eclipses by material in the disk. With $f$ of 0.11 as estimated above, we expect an opening angle of $\alpha = 20$\,deg. This number is currently based on a very small sample, and needs to be validated when larger numbers of these variable objects are available.

For comparison, the opening angle in moderately flared disks around low-mass T Tauri stars is about 5\,deg \citep{2016A&A...594A..83D}). Our targets have evolved disks which are probably flatter than those in younger stars, thus the opening angles should be significantly smaller than 5\,deg and much smaller than 20\,deg. In a sample of 17 stars with disks at an opening angle of 5\,deg, we would expect 0.5 to be in the line of sight -- but we find 2. This tentatively means that one of the assumptions made above has to be wrong. It is contrived to argue that inclinations are non-random. More plausible, the ring that is causing the eclipses may not be confined to a thin disk. It is not simply a part of the disk or as thin as the disk, but instead a separate structure.

\section{Discussion}

\subsection{Comparison with related stars}

Obscurations by circumstellar material are a common phenomenon among very young T Tauri stars with primordial disks. For low-mass stars, AA Tau has become the prototype for a class of 'dippers', accreting stars that are occulted by a warp in the inner disk, with events repeating on timescales of days, presumably caused by the interaction of the inner disk with the stellar magnetic field \citep{2007A&A...463.1017B,2015AJ....149..130S}. About 10-20\% of all young stars belong to that category \citep{2015A&A...577A..11M}. Once accretion ceases and the disks become passive, this phenomenon is expected to subside. Therefore it is puzzling to find 'dipping' behaviour, particularly without regular pattern, among significantly older stars with evolved disks, like our four targets. 

There are, however, other examples of stars at comparable age and mass that also show 'dips' best explained by circumstellar matter travelling through the line of sight, and it is instructive to compare with those. In Table \ref{table3} we summarise the properties of the four stars in $\epsilon$\,Ori and four selected comparison objects. J1407 is discussed in the literature as a star periodically eclipsed by a giant ring system around a substellar companion \citep{2012AJ....143...72M,2015ApJ...800..126K}. It exhibits huge, long eclipses, with a regular pattern and on timescales of years. PDS110 could be a similar system at earlier spectral type with some remaining IR excess, although the repetitive nature has not been confirmed yet \citep{2017MNRAS.471..740O}. These two are clearly of a different nature than our four examples in terms of their phenomenology and their physical interpretation.

RZ\,Psc's eclipses are shorter, over a couple of days, similar to our targets. The star also has a spectral energy distribution that resembles our targets. However, the eclipses recur over much longer timescales, and most of the time the star is close to its upper brightness level. The variability is explained as the result of an asteroidal ring \citep{2017RSOS....460652K} oriented in the line of sight. This is similar to our interpretation for the $\epsilon$\,Ori stars, except that we argue for a vertically extended ring in addition to the flat disk (see Sect. \ref{sed}), and it is unclear if the same applies to RZ\,Psc's ring. RZ\,Psc might be surrounded by a more dispersed, low-density version of the rings we expect to be present in our targets -- perhaps fitting, since RZ\,Psc is also somewhat older and more evolved. 

RIK-210 is a non-accreting low-mass star in the Upper Scorpius association undergoing dips of variable depth and morphology, with marginal excess at 22$\,\mu m$, not unlike those observed for debris disks \citep{2017ApJ...835..168D}. The discovery paper deems eclipses by circumstellar matter as 'extremely unlikely' due to the lack of IR excess, but as we have shown in Sect. \ref{lc_ana}, this does not necessarily preclude the existence of the amount of dust needed for such events. The eclipses are typically 10\% deep and a few hours long. While there are many similarities with our targets in the $\epsilon$\,Ori region, including age and mass, the main difference seems to be that the dips in RIK-210 occur at a regular period of 5.7\,d, in phase with the rotation period, which is also seen in the lightcurve as spot modulation. No period is observed for our stars. 

A number of non-accreting stars with lightcurve dips -- all in Upper Scorpius and all very low mass stars -- have recently been found using lightcurves from the Kepler/K2 mission \citep{2016ApJ...816...69A,2017AJ....153..152S,2018AJ....155...63S}. All of them show flux dips in Kepler/K2 data, sometimes synchronous with the rotation period, with a wide range in dip duration and depth. A few of them are similar in their lightcurve characteristics to the previously mentioned RIK-210 \citep{2017AJ....153..152S}, whereas others are more reminiscent of AA-Tau-like dippers \citep{2016ApJ...816...69A}. In terms of the dip depth and duration, two of our four stars (V1999 and V1959) are comparable to these stars in Upper Scorpius. V2227 has faster dips than any objects in the literature, whereas V2559's dips are much deeper. These two are clearly exceptional objects.

Whereas the objects discussed by \citep{2017AJ....153..152S} mostly do not show IR excess at wavelengths up to 12$\,\mu m$ (i.e. at most they host a very evolved or debris disk), the ones shown by \cite{2016ApJ...816...69A} have full or evolved disks. This largely drives the interpretation given in these papers. Ansdell et al. explain the dips with transiting circumstellar dust clumps, a scenario similar to our interpretation of the $\epsilon$\,Ori variables. For objects with resolved submm images, it is shown that the disks are not close to edge-on, i.e. the clumps do not reside in the disk \citep{2016MNRAS.462L.101A}. Stauffer et al. prefer an explanation related to magnetic activity, the presence of warm coronal clouds trapped in prominences. We would like to point out, however, that the presence of dusty rings is not strictly ruled out by their data. 

In summary, based on their lightcurve morphology, the four stars in $\epsilon$\,Ori belong to a growing category of post-T Tauri, non-accreting stars with variability caused by structures in their circumstellar environment along the line of sight. For many of them, dusty rings are a plausible explanation. The characteristics of the dips seem to show a relation to the presence of IR excess -- stars with detectable excess in the mid-infrared caused by a disk (like V1959, V1999, V2559, or the 'dippers' by Ansdell et al.) show deep and often irregular dips, those without IR excess (which might still have small amounts of circumstellar dust) have more regular, shallow dips. 

This could point to an evolutionary sequence: At about the time when accretion ceases and the primordial disk begins to flatten and dissipate, stars develop inhomogenuous rings that contain a number of dusty clouds. The rings may be transient and fed by the still existing dust reservoir in the disk, or they may be made of gravitationally bound objects. Over time, the rings dissipate and leave behind only discrete clouds that cause more regular dips. This would explain why this phenomenology is mostly observed in a limited age range roughly between 5 and 10\,Myr, a phase in which the mass of the dust in the disks drops quickly from about a hundred to a fraction of an Earth mass \citep{2013MNRAS.435.1037P} and the disks change from primordial dust to debris. Dusty rings might be a common feature associated with this fundamental transition. 

\begin{table*}
\caption{Summary of observational features for selected post-T Tauri variables with partially irregular eclipses. $\Delta t$ is the duration of the typical eclipse, $P$ their period (if there is any) or typical separation between minima. IR indicates the presence of IR excess at $<5\,\mu m$ (NIR) or $>5\,\mu m$ (MIR). }
\label{table3}
\begin{tabular}{|llllllll|}
\hline
object   & SpT  & $\Delta m$& $\Delta t$ & $P$     & NIR & MIR  & references\\	       
\hline
V1959    & M3   & $>0.05$       & 5-10\,h    & 1-2\,d  & yes & yes  & this paper\\
V1999    & M4   & $>0.2$	    & 5-10\,h    & 1-2\,d  & yes & yes  & this paper\\
V2227    & M3   & $>0.2$	    & 0-1\,h     & hours   & no  & ?    & this paper\\
V2559    & M3   & $>1.0$	    & 4-?\,d     & days	   & no  & yes  & this paper\\
\hline
RZ Psc   & K0   & $<2.5$    & days       & 70\,d?  & yes & yes  & \citet{2017RSOS....460652K}  \\
RIK-20	 & M2.5 & $0.1-0.2$ & 0.2-0.8\.d & 5.7\,d  & no  & no?  & \citet{2017ApJ...835..168D} \\
J1407    & K5   & $1-3$     & 56\,d      & years   & no  & no?  & \citet{2012AJ....143...72M} \\
PDS110   & F6   & 0.25      & 25\,d      & 808\,d? & yes & yes  & \citet{2017MNRAS.471..740O}\\
\hline
\end{tabular}
\end{table*}

\subsection{Origin of vertically extended rings}

The formation of large-scale rings is one of the features in the late evolution of accretion disks \citep{2007MNRAS.375..500A} and in the transition from primordial to debris disks \citep{2015Ap&SS.357..103W}. Some of these rings ('exozodiacal') are thought to be located very close to the central stars, as evidenced by near/mid-infrared flux excess, see the recent review by \citet{2018arXiv180204313H}. The ALMA pictures of protoplanetary disks demonstrate that broad rings in disks are common at much earlier evolutionary stages, too \citep{2015ApJ...808L...3A,2018ApJ...869...17L}. These previously observed structures, however, are all confined to a relatively thin disk, and thus would not be expected to cause variability in a significant fraction of stars -- in contrast to the phenomena discussed in this paper. For the stars discussed here, a mechanism is needed to scatter the occulting objects to large scale heights. In gas-dominated primordial disks, plausible ways to lift dust out of the disk include magnetospheric interaction between star and disk or disk winds, but in non-accreting stars these processes are not expected to be present anymore. 

Therefore we have to look elsewhere to explain the rings that are observed through the induced variability. Given the ubiquity of planetary systems in field stars, it is likely that planets or planetesimals reside in many disks. Planet forming processes, planet migration, and the interaction of planets with their environments are expected to create a variety of structures and asymmetries within the disks \citep{2017arXiv170308560K}. 

If the clouds that constitute the rings are gravitationally bound to central objects (e.g., planets, planetesimals, comets, asteroids), they could be scattered out of the disk by direct interactions with other objects or disk structures. Alternatively, if the clouds are tenuous without central object, they could be caused by a migrating planet in the disk that stirs up a population of planetesimals and creates a small amount of dust at high inclinations, as simulated by \citet{2011A&A...531A..80K} for Neptune-mass planets. A more detailed discussion of these scenarios is beyond the scope of this paper and left for future work. In any case, the eclipses observed here are possible signposts for the presence of a young, evolving planetary system and for a dynamic phase in the planet-disk interaction. These stars are therefore excellent targets for future searches for young planets.

\section*{Acknowledgements}

The processed data and scripts used for the analysis in this paper are available under \url{https://notebooks.azure.com/aleksthethird/projects/dippers}. 

This publication makes use of data products from a variety of missions, including WISE, NEOWISE (both projects of JPL/Caltech, funded by NASA), SDSS (funded by the Alfred P. Sloan Foundation, the Participating Institutions, the NSF, and the U.S. Department of Energy Office of Science), 2MASS (a joint project of the University of Massachusetts and IPAC/Caltech, funded by NASA and NSF), CRTS \citep{2009ApJ...696..870D} and the ESA mission Gaia (\url{https://www.cosmos.esa.int/gaia}). 
Our own observations were carried out with the LCOGT network using observing resources funded by the Scottish Universities Physics Alliance. For data reduction and analysis, we used matplotlib \citep{2007CSE.....9...90H}, NumPy \citep{2011arXiv1102.1523V}, and IRAF \citep{1993ASPC...52..173T}. For data retrieval, we made use of the SIMBAD database and the VizieR catalogue access tool, both operated at CDS, Strasbourg, France, TOPCAT \citep{2005ASPC..347...29T}, as well as the NASA/IPAC Infrared Science Archive, operated by the JPL/Caltech under contract with NASA. The compilation of acknowledgements was helped by the Astronomy Acknowledgement Generator.  

This project was supported by STFC grant ST/R000824/1 to AS. MP acknowledges funding from the European Research Council (ERC) under the European Union's Horizon 2020 research and innovation programme via the ERC Starting Grant MUSTANG (grant agreement number 714907), and from the ERC under ERC-2011-ADG via the ECOGAL project (grant agreement number 291227). We are grateful to the staff at the Dublin Institute for Advanced Studies for their hospitality and generosity which facilitated the research for this paper.









\bsp	
\label{lastpage}
\end{document}